\documentclass[amsmath,amssymb,a4paper,11pt]{article}
\usepackage{graphicx,wrapfig}% Include figure files
\usepackage{bm}% bold math
\usepackage{amsmath}
\usepackage{amssymb}
\usepackage{url}
\usepackage{cite}
\usepackage{subcaption}
\usepackage{stmaryrd}
\usepackage[capitalize]{cleveref}
\usepackage{color}
\usepackage{enumitem}
\usepackage{epstopdf}
\usepackage[normalem]{ulem}
\newcommand{\rb}{\boldsymbol{r}}
\newcommand{\pb}{\boldsymbol{p}}
\newcommand{\qb}{\boldsymbol{q}}
\newcommand{\ab}{\boldsymbol{a}}

\newcommand{\vb}{\boldsymbol{v}}

\newcommand{\kb}{\boldsymbol{k}}

\newcommand{\Rb}{\boldsymbol{R}}

\newcommand{\ep}{\varepsilon}

\newcommand{\sumprime}[1]{\sum_{#1}{\vphantom{\sum}}^{\!\!\prime}}

\begin{document}
\begin{center}
{\Large\bf Self-diffusion in a spatially modulated system of electrons on helium}

{\large{\bf K. Moskovtsev}\footnote{e-mail:kmoskovtsev@gmail.com} and  {\bf  M. I. Dykman}\footnote{e-mail:dykman@pa.msu.edu}}

{\it Department of Physics and Astronomy, Michigan State University, East Lansing, MI 48824, USA}

\end{center}

\noindent
{\bf Abstract.} {\small  We present results of molecular dynamics simulations of the electron system on the surface of liquid helium. The simulations are done for 1600 electrons with periodic boundary conditions. Electron scattering by capillary waves and phonons in helium is explicitly taken into account. We find that the self-diffusion coefficient superlinearly decreases with the decreasing temperature. In the free electron system it turns to zero essentially discontinuously, which we associate with the liquid to solid transition. In contrast, when the system is placed in the fully commensurate one-dimensional potential the freezing of the diffusion occurs smoothly. We relate this change to the fact that, as we show,  a Wigner crystal in such a potential is stable, in contrast to systems with a short-range inter-particle coupling. We find that the freezing temperature nonmonotonically depends on the commensurability parameter. We also find incommensurability solitons in the solid phase. The results reveal peculiar features of the dynamics of a strongly correlated system with long-range coupling placed into a periodic potential.}

\hfill

%\noindent
%\today

%%%%%%%%%%%%%%%%%%%%%%%%%%%%%%%%%%%%%%%%%%%%%%%%%%%%%%%%%%%%%%%%%%%%%%%%%
%%%%%%%%%%%%%%%%%%%%%%%%%%%%%%%%%%%%%%%%%%%%%%%%%%%%%%%%%%%%%%%%%%%%%%%%%
\section{Introduction}

Electrons on helium form a strongly correlated nondegenerate liquid or a Wigner crystal. The profound effects of electron correlations have been studied at length theoretically and experimentally \cite{Andrei1997,Monarkha2004}. However, the detailed dynamics of the electrons in the liquid has been explored only indirectly, through its effects on the behavior at the wavelengths long compared to the interlectron distance, such as magnetotransport, tunneling, and a number of resonant nonlinear phenomena \cite{Menna1993,Dykman1993i,Lea1997,Dykman2001a,Konstantinov2009,Konstantinov2013,Chepelianskii2015}. An insight into the short-wavelength dynamics has been coming from modeling the system, and a significant number of molecular dynamics simulations and Monte Carlo simulations have been done over the last four decades, cf. \cite{Totsuji1978,Gann1979,Hansen1979,Morf1979,Kalia1981,Kalia1983,Strandburg1988,FangYen1997,Muto1999,Piacente2005,Clark2009,damasceno2010, Rees2012,Mazars2015,Khrapak2018}. Much of the emphasis was placed on the melting transitions and the occurrence of the hexatic phase. The recent Monte Carlo results for a record large system indicate that the hexatic phase emerges, but only in a vary narrow range of temperatures \cite{Mazars2015}. An advantageous feature of molecular dynamics simulations is that they allow one to see how the electron system evolves in time. However, so far such simulations were done using either Langevin equations of motion with a phenomenologically added friction force, which does not describe the electron dynamics on helium, or using a comparatively short thermalization time, which may be insufficient, at least in the transition region (for example, in the Monte Carlo simulations \cite{Mazars2015} a much longer equilibration time was used). 

In this paper we develop a numerical algorithm that allows one to take into account  the actual microscopic mechanism of the electron scattering off the excitations in helium along with the long-range electron-electron interaction. Using this algorithm, we perform molecular dynamics simulations of a large electron system comprised of 1600 electrons with periodic boundary conditions. To the best of our knowledge, this is the largest system used in molecular dynamics simulations of two-dimensional electron systems. An important characteristic of the electron dynamics is the coefficient of self-diffusion. It describes the motion of individual electrons on a mesoscopic scale that exceeds the interelectron spacing, but is small on the macroscopic scale. We study self-diffusion in the presence of electron scattering by helium excitations. It should be emphasized that the coefficient of self-diffusion is not described by the standard Einstein relation between diffusion and mobility, which refers to the macroscopic diffusion and the long-wavelength mobility.  

A characteristic feature of the electron system is that it remains a liquid where the ratio $\Gamma$ of the electron-electron interaction energy $E_C$ to the kinetic energy is very large, 
\begin{align}
\label{eq:Gamma}
\Gamma = E_C/k_BT, \qquad E_C=e^2(\pi n_s)^{1/2}.
\end{align}
Here, $n_s$ is the electron density and $T$ is temperature. For $\Gamma\gg 1$ self-diffusion should reflect a correlated motion in the system. Diffusive motion should largely stop once the system crystallizes, and therefore the self-diffusion coefficient is an important indicator of crystallization. We find that it goes to zero  extremely sharply which, we believe, is also a consequence of the strong electron correlations. The peculiar features of the transition are due to the long-range electron-electron interaction and the absence of long-wavelength longitudinal acoustic phonons in the electron crystal. 

Electrons on helium is a unique condensed-matter system in that it allows studying many-body phenomena that are not masked by disorder due to impurities and other defects present in solid-state systems. An important group of these phenomena are related to commensurate-incommensurate transitions \cite{Bak1982}. They occur where a crystalline monolayer is placed on a crystalline substrate with close lattice spacing \cite{Pokrovskii1980}. The commensurate-incommensurate transitions occur also in a macroscopic system of colloidal particles in a periodic potential created by laser radiation \cite{Wei1998,Radzihovsky2001}. The motion of colloidal particles is usually overdamped, and therefore the dynamics is qualitatively different from that of solid-state systems.

For electrons on helium, a tunable periodic potential can be created by placing a periodic structure beneath the helium surface. Most interesting effects are expected to occur for a structure with a period on the order of the interelectron distance. Since this distance is $\sim 1\,\mu$m, such a structure can be made using standard fabrication techniques. A periodic potential can significantly, and nontrivially, affect not only the electron solid, but also a strongly correlated electron liquid, and in particular the dynamics of the electron liquid. This should happen just because the spatial structure of the liquid should be sensitive to a periodic potential where the period is close to the interelectron distance. Self-diffusion is a natural characteristic of the changes in the dynamics. The first molecular dynamics simulations of self-diffusion in a periodic potential  were described in the interesting paper \cite{Kalia1983}. The results were obtained for a system of 256 electrons, with no coupling to helium excitations, for $\Gamma=36$ and for several values of the integer and half-integer ratio of the distance between the electron rows in the Wigner crystal and the modulation period.  

In this paper we study self-diffusion and freezing of the electron system on helium placed in a one-dimensional (1D) potential. The potential slows down the dynamics, and  we found that obtaining reliable statistics requires very long simulations. We find that the electron thermalization improves when scattering by the helium excitations is taken into account. Our simulations allow us  to characterize the anisotropy of self-diffusion close to the commensurability of the periodic potential, including the detailed dependence of the self-diffusion coefficients on the control parameter $\Gamma$.

In Sec.~2  we describe the model and the algorithm of the numerical integration that takes into account elastic and inelastic scattering of electrons on the helium surface. In Sec.~3 we describe self-diffusion and its sharp change at the freezing transition in the absence of a periodic potential. We also describe the short-wavelength structure factor. In Sec. 4 we study self-diffusion in an external 1D periodic potential, which is maximally commensurate with the electron crystal. We show that, for a Wigner crystal in such a potential, the mean-square electron displacement from the lattice site does not diverge, which leads to a qualitative change of the liquid to solid transition. We show that the diffusion becomes strongly anisotropic and study its dependence on $\Gamma$ and the amplitude of the periodic potential. In Sec. 5 we study the diffusion in the system where the potential is close to maximal commensurability. We also explore the possiblity of the onset of a lattice of solitons in the electron density close to maximal commensurability. Section 6 contains concluding remarks.

%%%%%%%%%%%%%%%%%%%%%%%%%%%%%%%%%%%%%%%%%%%%%%%%%%%%%%%%%%%%%%%%%
%%%%%%%%%%%%%%%%%%%%%%%%%%%%%%%%%%%%%%%%%%%%%%%%%%%%%%%%%%%%%%%%%
\section{Model of the electron system and the numerical approach}
\label{sec:model}

\subsection{The Hamiltonian}

We consider a two-dimensional (2D) many-electron system with the Coulomb electron-electron interaction. The system is placed into an external one-dimensional periodic potential $U(x)$. The electrons are interacting with surface capillary waves on helium, ripplons, and with phonons in helium. The Hamiltonian is
\begin{align}
\label{eq:Hamiltonian}
&H=H_{ee} + H_U + H_{\mathrm He} + H_i, \quad H_{ee} = \sum_n\frac{\pb_n^2}{2m} + \frac{1}{2}\sum_{n\neq m}\frac{e^2}{|\rb_n-\rb_m|},\nonumber\\
&H_U = \sum_n U(\rb_n), \quad H_{\mathrm He} = \hbar\sum_{\qb,q_z}\omega_{\qb,q_z}a^\dagger_{\qb,q_z} a_{\qb,q_z} +
\hbar\sum_{\qb}\omega_{\qb} b^\dagger_{\qb} b_{\qb}
\end{align}
Here, $\rb_n$ and $\pb_n$ are the 2D coordinate and momentum of an $n$th electron, $U(\rb)$ is the external potential, $a_{\qb,q_z}$ is the annihilation operator of a phonon in helium with 2D wave vector $\qb$ and the wave number $q_z>0$ of motion transverse to the surface, $\omega_{\qb,q_z}$ is the phonon frequency, $b_{\qb}$ is the ripplon annihilation operator, and $\omega_{\qb}$ is the ripplon frequency. To calculate the electron-phonon coupling, one should keep in mind that the density modulation $\delta\rho_{\rm He}$ in the phonon field as a function of the in-plane coordinate $\rb$ and the coordinate $z$ normal to the surface is 
\[\delta\rho_{\rm He}(\rb, z) =-i\sum_{\qb,q_z} (\hbar \rho_{\rm He}\omega_{\qb,q_z}/Vv_{\rm He}^2)^{1/2}e^{i\qb\rb} \sin (q_z z)(a_{\qb,q_z}- a^\dagger_{-\qb,qz}),\]
where $V$ is the helium volume and $v_{\rm He}$ is the sound velocity.

Ripplons are very soft excitations. For typical wave vectors of the ripplons that scatter electrons $\qb \sim \pb/\hbar$ the ripplon energy $\hbar\omega_{\qb}$ is much smaller than the electron energy. Therefore scattering by ripplons is essentially elastic. Inelastic scattering comes from two-ripplon processes and from scattering by phonons. Generally, these two mechanisms give comparable scattering rates, with the phonon scattering rate being slightly higher. However, the overall rate of inelastic scattering is orders of magnitude smaller than the rate of elastic scattering, for thermal electron energies \cite{Andrei1997}. In our analysis the role of inelastic scattering is to make sure the electron energy distribution remains close to equilibrium, and therefore it suffices to take into account just one mechanism of inelastic scattering, which we chose to be the scattering by phonons. Respectively, we consider one-ripplon and one-phonon coupling,
\begin{align}
\label{eq:coupling}
H_i = \sum_n\sum_{\qb}V_{\qb}^{\mathrm {(rp)}}(z_n)e^{i\qb\rb_n}(b_{\qb}+b_{-\qb}^\dagger) + 
\sum_n\sum_{\qb,q_z}V_{\qb,q_z}^{\mathrm {(ph)}}(z_n)e^{i\qb\rb_n}(a_{\qb,q_z}- a^\dagger_{-\qb,q_z})
\end{align}
The parameters $V_{\qb}^{\mathrm {(rp)}}(z)$ and $V_{\qb,q_z}^{\mathrm {(ph)}}(z)$ of the coupling to ripplons and to phonons as functions of the distance $z>0$ of the electron from the helium surface are well-known \cite{Andrei1997,Dykman2003,Schuster2010a}. We will be interested in the case of a weak field that presses electrons to the surface. Respectively, we keep in $H_i$ only the terms related to the change of the image potential by the vibrational excitations; moreover, in the case of the coupling to phonons, we keep only the terms related to the phonon-induced spatial modulation of the helium dielectric constant [which is $\propto \delta \rho_{\rm He}(\rb,z)$]; the effect of the phonon-induced change of the shape of the helium surface is comparable, but somewhat smaller. Then  
\begin{align}
\label{eq:matrix_elements}
&V_{\qb}^{\mathrm {(rp)}}(z)=\Lambda\frac{(\hbar q)^{1/2}}{(2\rho\omega_{\qb}S)^{1/2}}\left\{z^{-2}[1-qz K_1(qz)]\right\}_z,\nonumber\\
&V_{\qb,q_z}^{\mathrm {(ph)}}(z)=-i\Lambda q(\hbar\omega_{\qb,q_z}/V v_{\rm He}^2\rho_{\rm He})^{1/2}\int_{-\infty}^0 dz'\frac{\sin (q_z z')}{z-z'}K_1\bigl(q(z-z')\bigr)
\end{align}
Here $\Lambda = e^2(\epsilon -1)/4(\epsilon+1)$ ($\epsilon$ is the dielectric constant of helium, $\epsilon\approx 1.057$) and $K_1(x)$ is the Bessel function; the expression for $V^{\mathrm (ph)}_{\qb,q_z}(z)$ corrects the typos in the corresponding expression in Refs.~\cite{Dykman2003, Schuster2010a}.

\subsection{Many-electron dynamics}

Strong correlations in the electron system can affect the electron scattering by ripplons and phonons. However, this effect is small in the absence of a magnetic field provided the electron dynamics is classical \cite{Dykman1997a}. The typical duration of a scattering event (collision) $t_{\rm coll}$  is determined by the kinetic energy of an electron, $t_{\rm coll} \sim \hbar/k_BT$. The kinetic energy is not a good quantum number because of the electron-electron interaction. The uncertainty of the kinetic energy of an electron is equal to the uncertainty of its potential energy in the electric field $E_{\rm fl}$ created by other electrons. This uncertainty is $\sim eE_{\rm fl}\lambda_T$, where $\lambda_T = \hbar/(mk_BT)^{1/2}$ is the thermal wavelength and thus the uncertainty in the electron position. In a strongly correlated system one can estimate $E_{\rm fl}$ assuming that electrons form a crystal and $eE_{\rm fl}$ is  a characteristic force on an electron performing small-amplitude vibrations about its equilibrium position (only short-range order is essential for the estimate of $E_{\rm fl}$ \cite{FangYen1997}). This gives $E_{\rm fl}\sim (k_BT n_s^{3/2})^{1/2}$. 

One can disregard the effect of the electron-electron interaction on collisions with ripplons and phonons if the duration of a collision is small compared to the characteristic time over which the electron kinetic energy is changed by the electron-electron interaction and the uncertainty of the kinetic energy is small compared to $\hbar/t_{\mathrm coll}$. From the above estimate, both conditions are satisfied if 
\begin{align}
\label{eq:classical_dynamics}
\hbar\omega_p\ll k_BT, \qquad \omega_p = (2\pi e^2n_s^{3/2}/m)^{1/2}.
\end{align}
Parameter $\omega_p$ is the short-wavelength plasma frequency of the electron system. This is also the short-wavelength phonon frequency in a Wigner crystal. The condition (\ref{eq:classical_dynamics}) shows that the electron dynamics is classical. We note that this condition must be satisfied where Monte Carlo or molecular dynamics simulations are used to determine the temperature of the transition into a Wigner crystal. In fact, it limits the numerical results on the crystallization to densities $n_s\lesssim 5\times10^7$~cm$^{-2}$, which is often lower than the densities where the crystallization is observed in the experiment.  

In our simulations we kept the effective density fixed, which corresponds to keeping $\omega_p$ fixed; different values of the plasma parameter $\Gamma$ correspond to different values of the temperature. It should be noted that the ``dynamical classicality condition'' (\ref{eq:classical_dynamics}) differs from the conventional condition that the electron system is classical if the Fermi energy $\ep_F= \pi\hbar^2 n_s/2m $ is small compared to $k_BT$. Indeed, the latter condition does not take into account the effect of the electron-electron interaction. One can see that $\ep_F/k_BT = (\sqrt{\pi}/4)(\hbar\omega_p/k_BT)^2 \Gamma^{-1}$. Therefore in the strongly correlated regime the inequality (\ref{eq:classical_dynamics}) is far more restrictive than $\ep_F\ll k_BT$.

\subsubsection{Integrating equations of motion}
\label{sec:integrating}
We performed molecular dynamics-type simulations of the many-electron system assuming that collisions with ripplons and phonons are instantaneous. The collisions happen at random and are not correlated. They are characterized by the rate  $w(\kb \to \kb')$ of electron scattering from the state with a given wave vector $\kb$ into the state with the wave vector $\kb'$  [$w(\kb\to \kb')$ is the probability density  per unit time and per unit area in the $\kb'$-space]. This rate  is given by the Fermi golden rule. Scattering by ripplons is elastic, and for this process $|\kb'| = |\kb|$ and the scattering rate $w(\kb\to\kb')$ depends only on the angle between $\kb$ and $\kb'$. 

\begin{figure}[h]
	\centering
		\includegraphics[width =\textwidth]{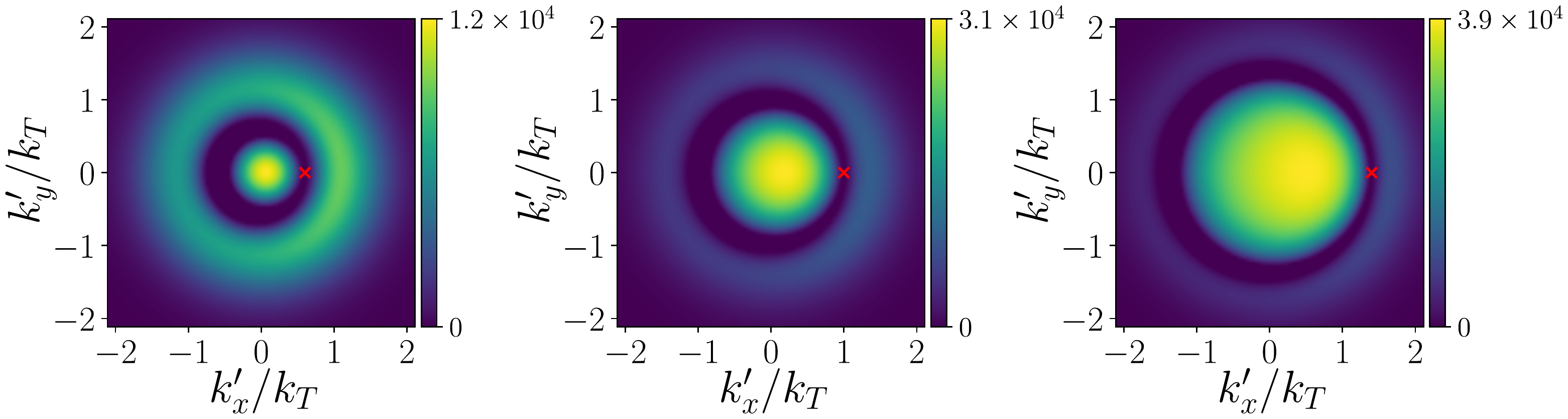}  
	\caption{\small The rate of scattering by phonons $w^{(\mathrm{ph})}(\kb\to \kb')$ as a function of $\kb'$. The left, central, and right panels refer to  $\kb/k_T = (0.6, 0)$, $\kb/k_T = (1, 0)$, and $\kb/k_T = (1.4, 0)$, respectively. Vector $\kb$ is shown by a red cross. The color shows $ k_T^2w^{(\mathrm{ph})}(\kb\to \kb')$ in s$^{-1}$. The scattering rate vanishes at points where $|\kb'| = |\kb|$. The plot refers to  $T=0.3$~K.}
	\label{fig:wkkp_phonon}
\end{figure}

Scattering by phonons is inelastic. The scattering rate $w^{\rm (ph)}(\kb\to\kb')$ is calculated on a grid in the space of the wave vectors $\kb$ and $\kb'$.  Importantly,  the change of the electron velocity in a collision $\hbar (\kb' - \kb)/m$ is large, of the order of the velocity itself. The rate $w^{\rm (ph)}(\kb\to\kb')$ as function of $\kb'$ is shown in Fig.~\ref{fig:wkkp_phonon} for several values of $\kb$. The calculation involves integration over the transverse to the surface component of the phonon wave vector. Function $w^{\rm (ph)}(\kb\to\kb')$  depends on temperature,we plot it by scaling $\kb$ and $\kb'$ by  the thermal wave number $k_T=(2mk_BT)^{1/2/}\hbar$.

The simulations are done by discretizing the time and assuming that, at equidistant instants $t_i$, there is a probability for an electron to change its velocity due to a collision with ripplons and phonons from $\hbar\kb/m$ to $\hbar\kb'/m$; it is given by  $(t_{i+1} - t_i)w(\kb \to \kb')d\kb'$. In the considered classical approximation, a collision is not accompanied by a change of the electron position. In the interval between $t_i$  and $t_{i+1}$ an electron is accelerated by the Coulomb force from other electrons. The difference in the position of an electron at instants $t_i$ and $t_{i+1}$ is calculated using the electron velocity at $t_i+0$ (i.e., with the account of the possible scattering at time $t_i$), which is incremented by the acceleration calculated for the positions of all electrons at time $t_i$ and multiplied by $(t_{i+1}-t_i)/2$. The discretized equations of motion are given in Appendix~\ref{sec:appA}. 

We consider electrons in the rectangular region with the size ratio $L_x/L_y=\sqrt{3}/2$ and impose periodic boundary conditions. When an electron leaves the main rectangle, it is injected on the other side of it. The geometry is chosen in such a way as to fit a triangular electron lattice into the main rectangle. Respectively, the mean inter-electron distance is $a_s =(2/n_s\sqrt{3})^{1/2} =L_y/N^{1/2}$, where $N$ is the number of electrons in the main rectangle and $N^{1/2}$ is an integer. Our simulations were carried out for $N=1600$. To allow for the long-range interaction, the  Coulomb force on an electron and the energy of the system are calculated using the Ewald summation in two dimensions, cf. \cite{Gann1979}. We tabulate the electric field from an electron and its periodically repeated images on a $2000\times 2000$ grid that covers the rectangle of size $(L_x/2,L_y/2)$, assuming that the electron is at the origin; the field at a general point is interpolated from the grid.

 We choose the external periodic potential $U(\rb)$ to be one-dimensional. This is the easiest form of a periodic potential that can be implemented in the experiment. It is constant along the $y$-coordinate, i.e., $U(\rb)\equiv U(x)$. To maintain the periodicity of the system, the period of $U(x)$ is set to be $L_x/M$ with an integer $M$.

%%%%%%%%%%%%%%%%%%%%%%%%%%%%%%%%%%%%%%%%%%%%%%%%%%%%%%%%%%%%

\section{Diffusion in a spatially uniform electron system}
\label{sec:uniform}

To calculate electron diffusion we keep track of the actual electron position in the extended system, that is, to find the electron displacement, the position after re-injection into the main rectangle is incremented by the length of the corresponding side of the rectangle. We calculate the displacement along the $x$ and $y$ axes separately. For a long calculation time of $10^7$ steps, the mean square displacement in the both directions is proportional to time. The proportionality coefficients are the coefficients of self-diffusion $D_x$ and $D_y$ along the shorter and the longer sides of the main rectangle, respectively.  Note that these are not 
the long-wavelength diffusion coefficients that describe the density current in response to a smooth electron density gradient. Therefore they are finite even in the absence of electron scattering by ripplons and phonons.

A natural scale for the self-diffusion coefficient in the electron liquid is
\begin{align}
\label{eq:diffusion_scale}
D_0=k_BT/m\omega_p \equiv \Gamma^{-1} (e^2/2m\sqrt{n_s})^{1/2} .
\end{align}
This value is obtained by taking into account that the liquid is strongly correlated for all values of the plasma parameter $\Gamma \gg 1$, and therefore it has a pronounced short-range order. The mean-square thermal  displacement of an electron about its quasi-equilibrium position in the liquid is of the same order of magnitude as if the electrons formed a crystal, $\sim k_BT/m\omega_p^2$, whereas the typical dynamical time in the strongly correlated system is $\omega_p^{-1}$, cf.  Eq.~(\ref{eq:classical_dynamics}). The estimate (\ref{eq:diffusion_scale}) is close to the De Gennes estimate \cite{Gennes1959} of the self-diffusion coefficient in liquids. Instead of $\omega_p$, De Gennes used a parameter of the same order of magnitude that depends on the pair correlation function of the liquid, which makes it temperature-dependent; however, the electron liquid that we study is quite different from a normal liquid, because it is two-dimensional and the electron-electron interaction is long-range; therefore we do not compare our results with the theory \cite{Gennes1959}. 

\begin{figure}[h]
\centering
\includegraphics[height=4.5truecm]{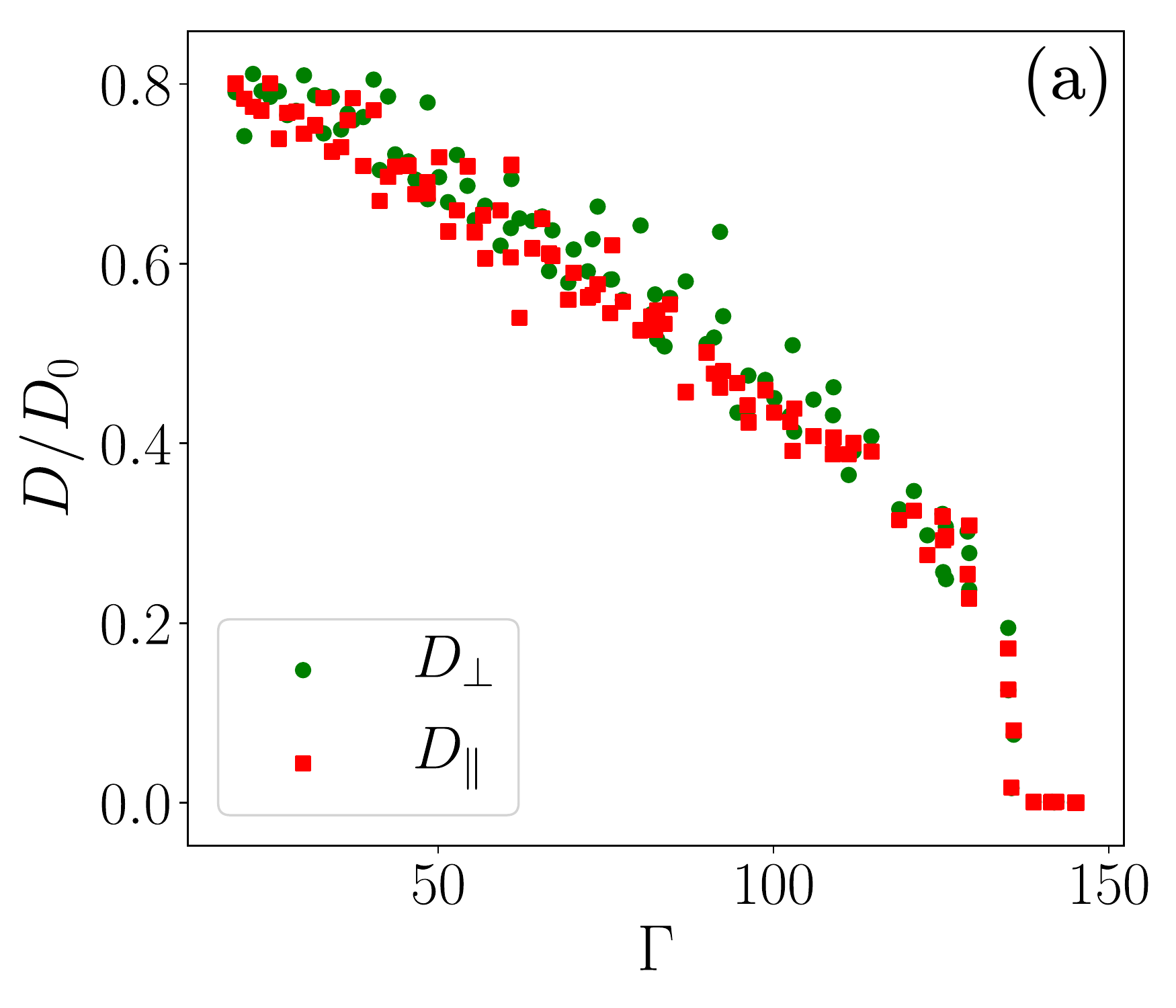}\hspace*{0.3in}
\includegraphics[height=4.5truecm]{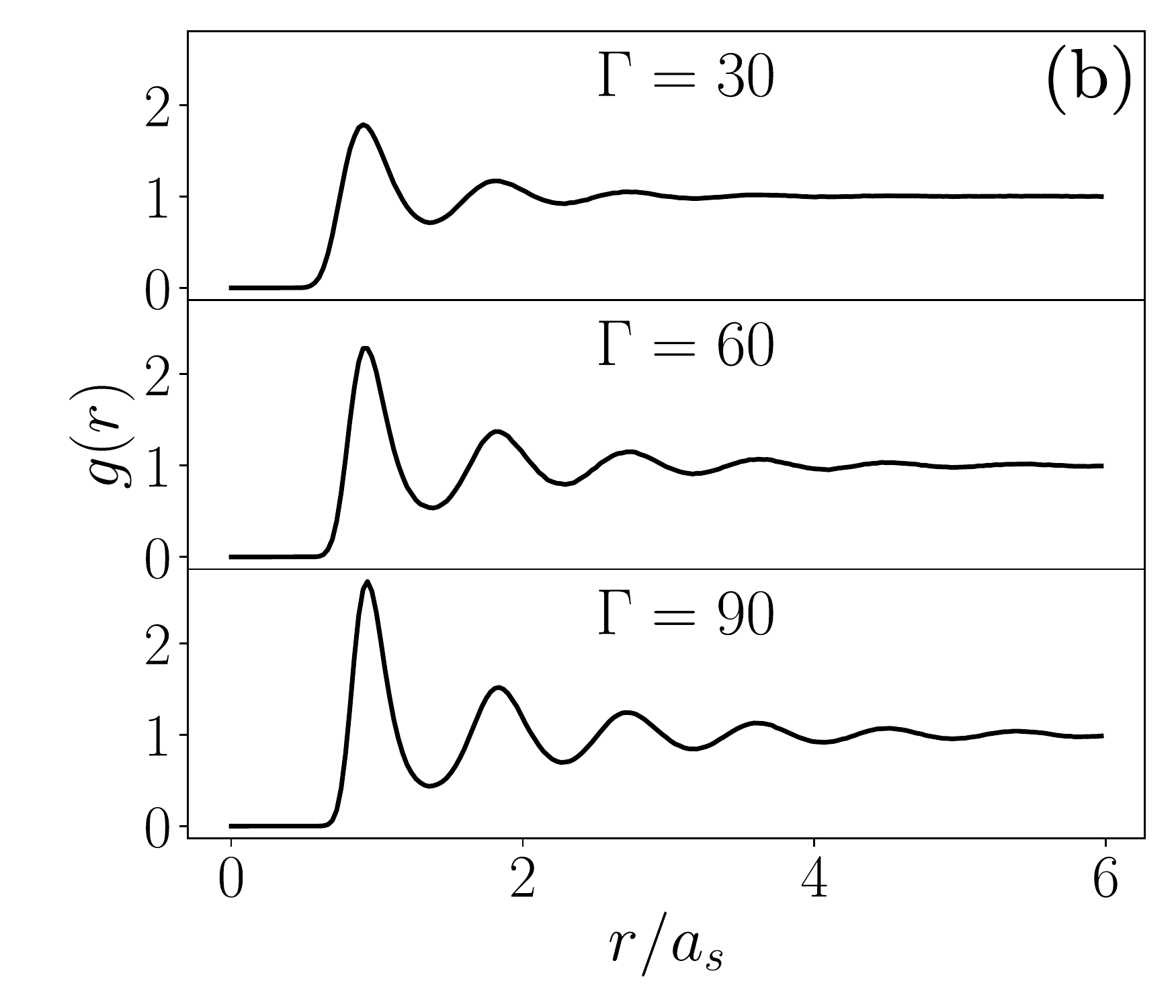}  \\
\includegraphics[height=4.5truecm]{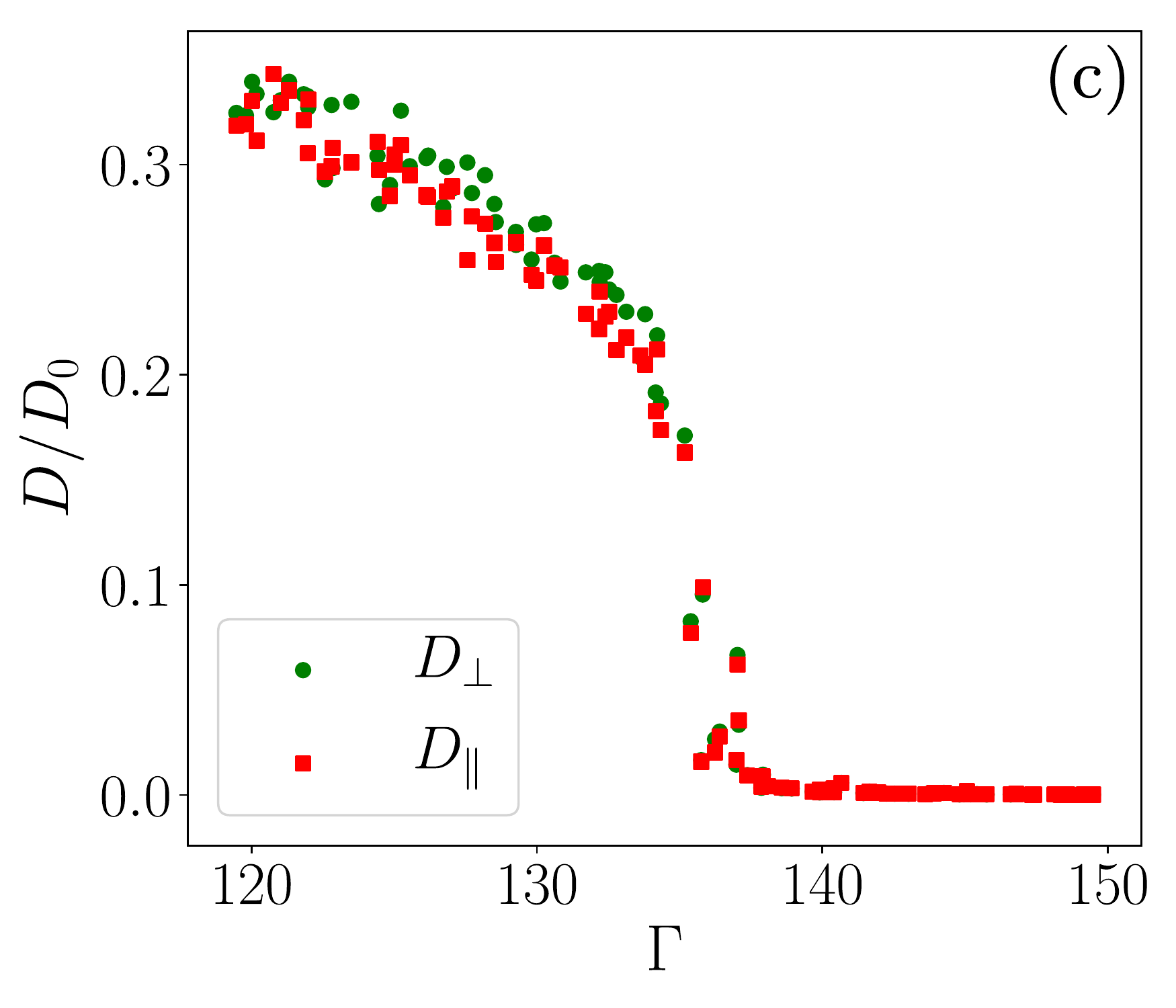} \qquad
\includegraphics[height=4.5truecm]{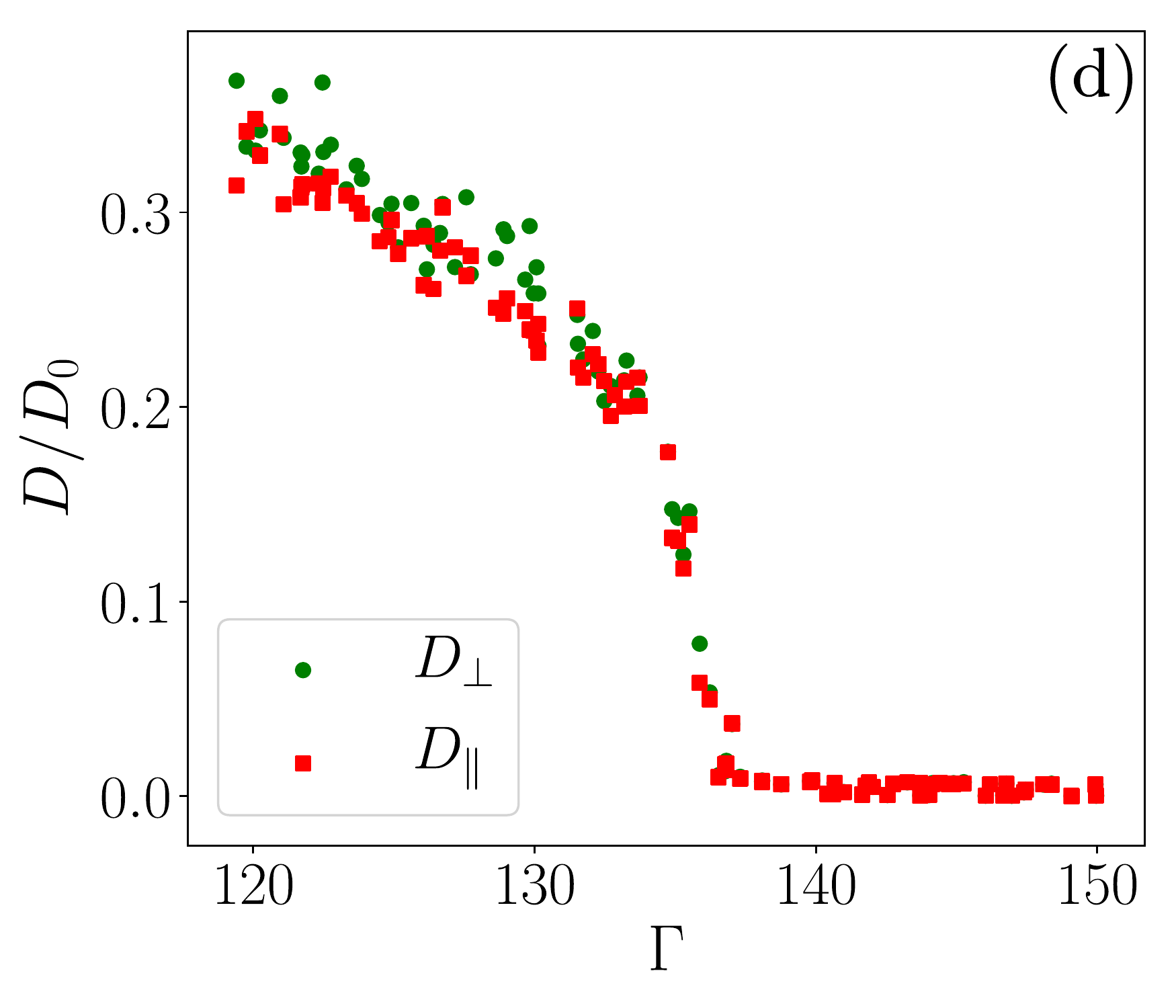}
	\caption{\small (a): The coefficients of self-diffusion $D_x$ and $D_y$ as functions of $\Gamma$. For each point $\Gamma$, the system is prepared as a triangular lattice that fits the main rectangle with the side ratio $L_x/L_y=\sqrt{3}/2$. The initial density is the same for all $\Gamma$, but the initial velocities are random. Then the system is equilibrated for $10^6$ time steps before collecting data for another $10^7$ steps. The simulation is carried out with the account taken of the both elastic and inelastic scattering by the excitations in helium. The electron temperature, and thus the value of $\Gamma$, is calculated after the equilibration period from the electron kinetic energy as $T=(m/2Nk_B)\sum {\bf v}_n^2$, where ${\bf v}_n$ is the velocity of the $n$th electron; it coincides with the temperature of phonons in helium. (b) The radial pair correlation function for the considered system of 1600 electrons with periodic boundary conditions. (c) The same as in (a), with higher resolution in the region of transition from the crystal to the liquid, (d) The diffusion coefficients in the transition region obtained by preparing the system initially in a spatially disordered state and letting the electrons come to equilibrium with the phonons at the chosen temperature, which requires $\lesssim 7\times 10^5$ steps. }
	\label{fig:diff_isotropic}
\end{figure}

The self-diffusion coefficients $D_x$ and $D_y$  are shown in Fig.~\ref{fig:diff_isotropic}(a) for a broad range of $\Gamma$. The finite amount of data used to average the squared displacements leads to an uncertainty in the coefficients $D_{x,y}$. The uncertainty is manifested as the spread of the data points in Fig. \ref{fig:diff_isotropic}(a). It decreases with the increasing system size or the simulation time.  In the absence of the periodic potential, $D_x$ and $D_y$ are close to each other, but $D_x$ is systematically slightly larger, which reflects the anisotropy imposed by the periodic boundary conditions with different lengths in the $x$ and $y$ directions. It is seen that both $D_x$ and $D_y$ remain on the order of $D_0$ and the ratios $D_x/D_0, D_y/D_0$ smoothly depend on $\Gamma$ in a broad range $25\lesssim \Gamma\lesssim 90$. 

With the increasing $\Gamma$, the correlation length in the system increases, cf. \cite{Gann1979}. This is seen in Fig.~\ref{fig:diff_isotropic}(b), which shows the radial distribution function
\[g(r) =[2\pi r n_s N(N-1)]^{-1}\sum_{n\neq m}\delta(r-|\rb_n-\rb_m|) 
.\]
The electron liquid becomes more ``rigid'' with the increasing correlation length. Respectively, the ratio $D_{x,y}/D_0$ decreases. Ultimately both $D_x$ and $D_y$ simultaneously sharply drop to zero, within the simulation precision. The transition to zero diffusion manifests crystallization of the system. The dependence of $D_{x,y}$ on $\Gamma$ in the transition region is shown in Fig.~\ref{fig:diff_isotropic}(c) and (d). We have found that the value of $\Gamma$ where the transition occurs depends on the size of the system and increases with the increasing size. This agrees with the conclusion drawn from the Monte Carlo simulations \cite{Mazars2015}. For 1600 electrons the transition region is $\Gamma \approx 137\pm 1$.

The results of two different Monte Carlo simulations led  to the conclusions that the transitions between the Wigner crystal and the hexatic phase is of the first order \cite{Clark2009} and the transition between the hexatic phase and the liquid is also a first order transition \cite{Mazars2015}. The sharp decrease of the diffusion coefficient in a narrow range of $\Gamma$ agrees with these conclusions. However, a comparison of Figs.~\ref{fig:diff_isotropic}(c) and (d) shows that there is no observable hysteresis when the system changes from the solid to the liquid phase and vice versa. This may be related to the uncertainty in the value of the diffusion coefficient and can be understood from the Monte Carlo result \cite{Mazars2015}, which suggests that the hexatic phase exists only in a very narrow range of $\Gamma$. The spread of the values of the diffusion coefficients does not allow us to resolve this range.

\begin{figure}[h]
	\centering
			\includegraphics[height=4.5truecm]{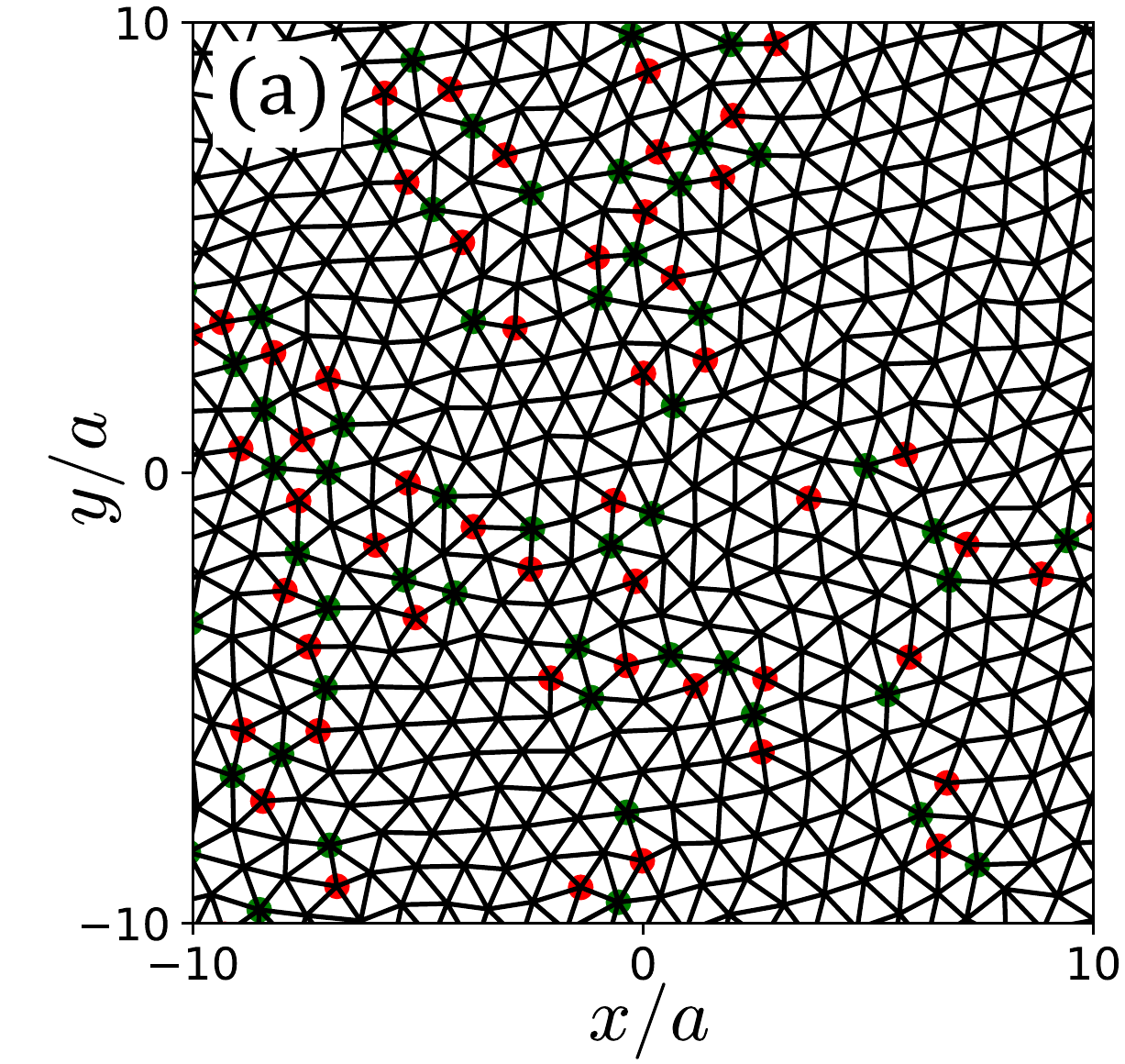} \qquad
			\includegraphics[height=4.5truecm]{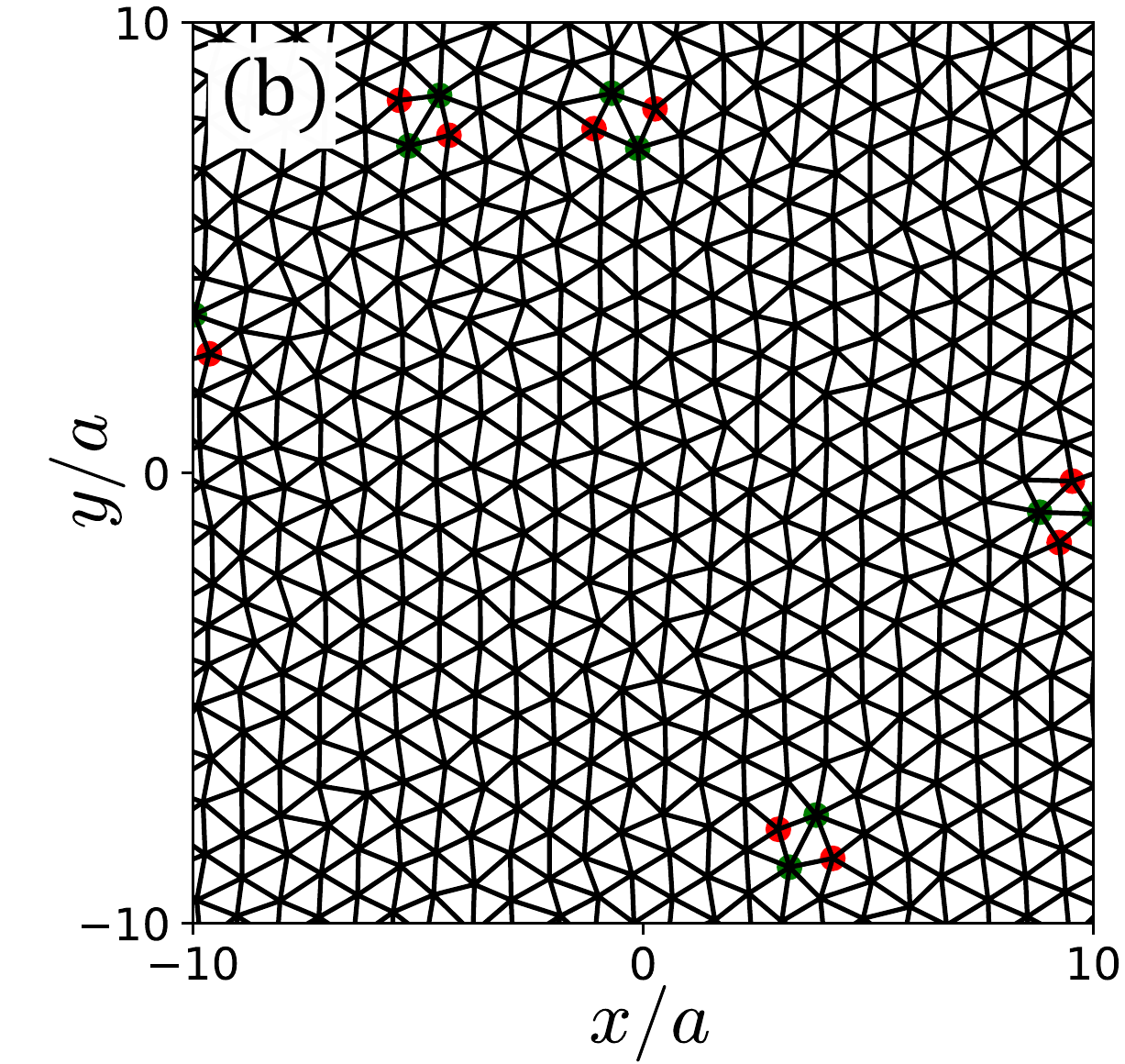}
	\caption{\small Delaunay triangulation of the electron positions for a portion of the system with $N=1600$ electrons in the presence of scattering by ripplons and phonons. Electrons are represented by vertices. The red vertices are electrons with five nearest neighbors, the green vertices are electrons with seven nearest neighbors. Pairs of red and green vertices represent dislocations. (a) Liquid phase with $\Gamma =133$ close to the freezing transition, the self-diffusion coefficients are $D_x \approx D_y\approx 0.2 D_0$. There are unbound dislocations,  pairs of bound dislocations, and unbound vortices with 5 or 7 nearest neighbors. (b) Solid phase with $\Gamma =140$ and $D_x = D_y = 0$. Only bound pairs of dislocations are present in the solid.}
	\label{fig:triangulation}
\end{figure}

The sharp increase of $D_{x,y}$ with decreasing $\Gamma$ in the region of the transition from the Wigner crystal to the electron liquid is a consequence of the sharp increase of the density of topological defects in this region. This is seen from Fig.~\ref{fig:triangulation}. The number of unbound dislocations and vortices, which correspond to isolated regions where an electron has more or less than 6 nearest neighbors, sharply increases with decreasing $\Gamma$ below the transition. When the density of these defects is small, the diffusion coefficient is approximately proportional to this density. When the density jumps up at the transition, so does also the diffusion coefficient. For somewhat smaller $\Gamma$, i.e., further away but not too far from the transition, the defect density is still low, but keeps increasing with the decreasing $\Gamma$. This is behind the superlinear increase of $D_{x,y}/D_0$ with the decreasing $\Gamma$ close to the transition. Once the density of defects is no longer small, the dependence of $D_{x,y}/D_0$ on $\Gamma$ becomes smooth. We remind that the simulations have been performed for constant electron density and the change of $\Gamma$ corresponds to the change of the temperature, with $D_0\propto T\propto \Gamma^{-1}$.

It should be noted that, sufficiently far from the region of the liquid to solid transition. the results on the diffusion coefficients in the presence of weak scattering by ripplons and phonons are very close to the results with no scattering However, in the transition region there is a difference. A discussion of this difference and its origin is beyond the scope of this paper.

%%%%%%%%%%%%%%%%%%%%%%%%%%%%%%%%%%%%%%%%%%%%%%%%%%%%%%%%%%%%

\section{One-dimensional commensurate periodic potential}
\label{sec:commensurte}

\subsection{The commensurability parameter}

In this and the following  sections we consider self-diffusion in the electron liquid placed into a periodic potential 
\begin{align}
\label{eq:potential}
U(x)=-A\cos Qx, \qquad Q=2\pi M/L_x,
\end{align}
where $M$ is an integer. Such form of a potential describes the potential from a periodic structure with period $2\pi/Q$ submerged beneath the helium surface by depth that exceeds $2\pi/Q$, in which case higher spatial harmonics of the potential can be safely disregarded. Parameter $A$ characterizes the strength of the modulation. A natural scale for $A$ is $k_BT$. The solid-state structure, which creates the potential $U(x)$ in the electron layer, modifies the electron-electron interaction,  because it leads to an image potential. If the structure is submerged by the depth of $4\pi/Q$, for example, and the voltage on the structure is 10~V, we have $A/k_B \approx 0.4$~K. The image potential depends on the details of the structure and, for a dielectric structure with a periodic set of interdigitized metal nanowires, the long-range Coulomb inter-electron potential will not be fully screened. As a result, as we discussed, the transitions between different phases of the electron system differ from those in system with short-range potential, cf.~\cite{Clark2009,Mazars2015}. To reveal the features of the electron dynamics on the helium surface related to the interplay of a periodic potential and the long-range electron-electron coupling, in this work we disregard the screening.

In our simulations the electron density $n_s$ is fixed, and then $k_BT$ is fully determined by the plasma parameter $\Gamma$, Eq.~(\ref{eq:Gamma}). Temperature is also explicitly contained in the strength of the coupling to ripplons and in the phonon distribution. However, for the considered weak coupling to ripplons and phonons, changing temperature just modifies the small scattering rate. For a given $\Gamma$, its effect on self-diffusion is minor; this should be contrasted with the change of the the long-wavelength mobility, which is limited by the scattering by ripplons and phonons. 

The electron dynamics sensitively depends on the interrelation between the period of the potential $2\pi/Q$ and the mean inter-electron distance $a_s$. If the electrons form a Wigner crystal that optimally fits into the main rectangle with the sides $(L_x,L_y)$, so that $L_y/a_s=2L_x/a_s\sqrt{3}$ is an integer, the distance between the neighboring electron rows in the $x$-direction is  $a_s\sqrt{3}/2$. One can then define the commensurability parameter \cite{Radzihovsky2001} as
\begin{align}
\label{eq:commensurability_parameter}
p_c = a_sQ\sqrt{3}/4\pi,
\end{align}
which is the ratio of the inter-row distance to the period of the potential. 
We will assume that the potential  $U(x)$ is weak compared to the potential of the electron-electron interaction, $A\ll e^2n_s^{1/2}$. The effect of the potential is most pronounced when $p_c$ is an integer or a fraction with small numerator and denominator, in which case the electron system can effectively adjust to the potential. In our simulations $p_c$ is always a fraction with not too large numerator and denominator. We find that, for a weak potential that we are studying, the ``incommensurability'' becomes strong already for $p_c \lesssim 0.3$. In what follows we use the term ``maximally commensurate'' for the case where $p_c=1$ and ``weakly incommensurate'' for $p_c \gtrsim 0.75$.

\subsection{Long-wavelength excitations in the solid phase}

In this section we consider  self-diffusion in the electron liquid for $p_c=1$, where the potential has a most pronounced effect. To gain intuition, we start with the analysis of the electron dynamics for $A\gg k_BT$ and $\Gamma\gg 1$ and assume that the electrons form a crystal. To minimize the energy, the crystal should be aligned with the potential, so that the electrons are at the troughs of $U(x)$.  It is convenient to describe the small-amplitude electron vibrations about the equilibrium positions $\Rb_n$ using the equations for the Fourier components $\delta\rb(\qb)$ of the electron displacements $\delta\rb_n=\rb_n - \Rb_n$. These equations have the form
\begin{align}
\label{eq:WC_commensurate}
\delta\ddot r_\alpha(\qb) = -\sum_\beta[\nu_0^2\delta_{\alpha\,x}\delta_{\alpha\beta} + C_{\alpha\beta}(\qb)]\delta r_\beta(\qb),\qquad \nu_0^2 = AQ^2 /m=8\pi^2n_sA/m\sqrt{3}. 
\end{align}
Here, the subscripts $\alpha,\beta$ enumerate the components $x,y$; $\nu_0$ is the frequency of vibrations of an isolated electron in the direction $x$ near the minimum of $U(x)$. Parameters $C_{\alpha\beta}$ describe the effect of the electron-electron interaction. For  small $q$ , $C_{\alpha\beta}(\qb)$ is a sum of the plasma contribution $C^{\rm (pl)}_{\alpha\beta}(\qb) = \omega_p^2 q_\alpha q_\beta/(qn_s^{1/2})$ and the contribution that describes the restoring force for transverse waves. For a hexagonal Wigner crystal the latter has the form \cite{Bonsall1977} $C^{\rm (tr)}_{xx}(\qb) = c_t^2 q_y^2 + a_{\rm BM}q_x^2$,  $C^{\rm (tr)}_{yy}(\qb) = c_t^2 q_x^2 + a_{\rm BM}q_y^2$,  and $C^{\rm (tr)}_{xy}(\qb) = (a_{\rm BM}-c_t^2)q_xq_y$. Here $c_t\approx 0.245e^2n_s^{1/4}/m$ is the transverse sound velocity in a Wigner crystal without an external periodic potential; $a_{\rm BM}\sim -5 c_t^2$.  

The confinement imposed by the periodic potential $U(x)$ changes the long-wavelength excitation spectrum. For $\omega_p q/n_s^{1/2}, c_t q\ll \nu_0$ the frequencies of the eigenmodes of the crystal are 
\begin{align}
\label{eq:frequencies}
\omega_1^2(\qb) \approx \nu_0^2 + \omega_p^2 q_x^2/(qn_s^{1/2}),\qquad \omega_2^2(\qb) \approx   \omega_p^2 q_y^2/(qn_s^{1/2}) + c_t^2 q_x^2 +a_{\rm BM}q_y^2
\end{align}
The second branch corresponds to the Goldstone mode in the crystal placed into a 1D commensurate potential with the displacement along the potential trough, $\omega_2(\qb)\to 0$ for $|\qb|\to 0$. The Landau-Peierls argument against the crystalline order in an infinite 2D system is based on the observation that the thermal mean-square displacement from a lattice site diverges. The divergence comes from the low-frequency acoustic phonons. From Eq.~(\ref{eq:frequencies}), the contribution of the low-frequency phonon branch (branch 2) to the mean-square electron displacement in the presence of a commensurate 1D periodic potential is
\begin{align}
\label{eq:displace_estimate}
\langle \delta\rb_n^2\rangle_2 \propto k_BT \int d\qb \omega_2^{-2}(\qb)=2\pi k_BT (n_s^{1/4}/\omega_p c_t)\int q^{-1/2} dq.
\end{align}
This integral converges at the lower limit $q\to 0$. Therefore one may expect that the commensurate potential stabilizes the crystalline configuration. In this case the nature of the transition from the electron liquid to the electron solid should change compared to the case of a free electron system.  

We should note that the estimate (\ref{eq:displace_estimate}) is specific for a Wigner crystal. The denominator in the right-hand side contains the characteristic parameter $\omega_p$ of the long-wavelength longitudinal plasma vibrations with frequency $\omega_p (q/\sqrt{n_s})^{1/2}$ for $q\to 0$ as well as the transverse sound velocity $c_t$. One therefore may expect that the liquid-to-solid transition will be different for a Wigner crystal from that in a crystal with short-range interaction placed into a 1D commensurate potential. 

\subsection{Electron diffusion and the liquid-to-solid transition}
\label{subsec:diffusion_commensurate}
 
\begin{figure}[h]
	\centering
			\includegraphics[width = 4.5truecm]{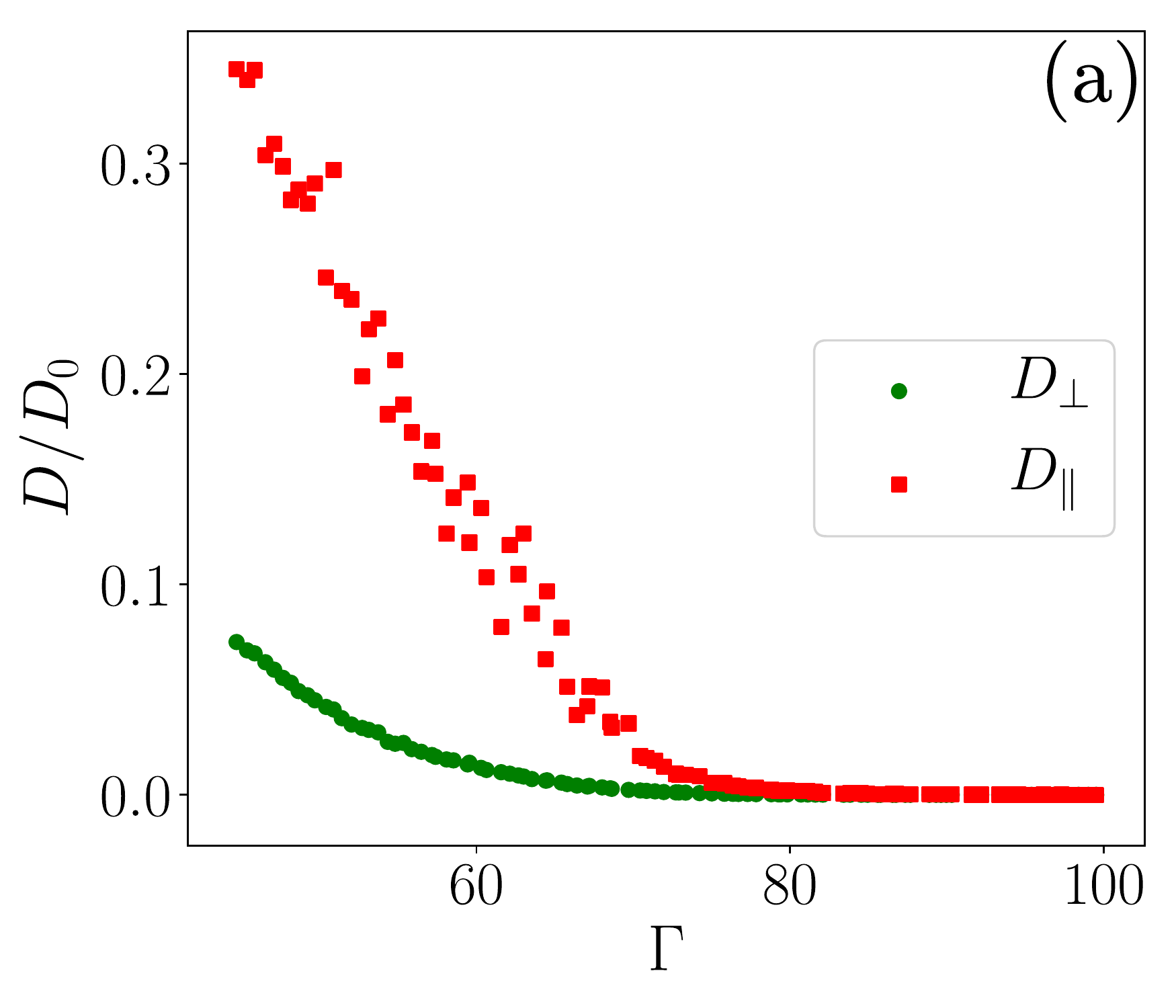} \qquad
			\includegraphics[width =4.5truecm]{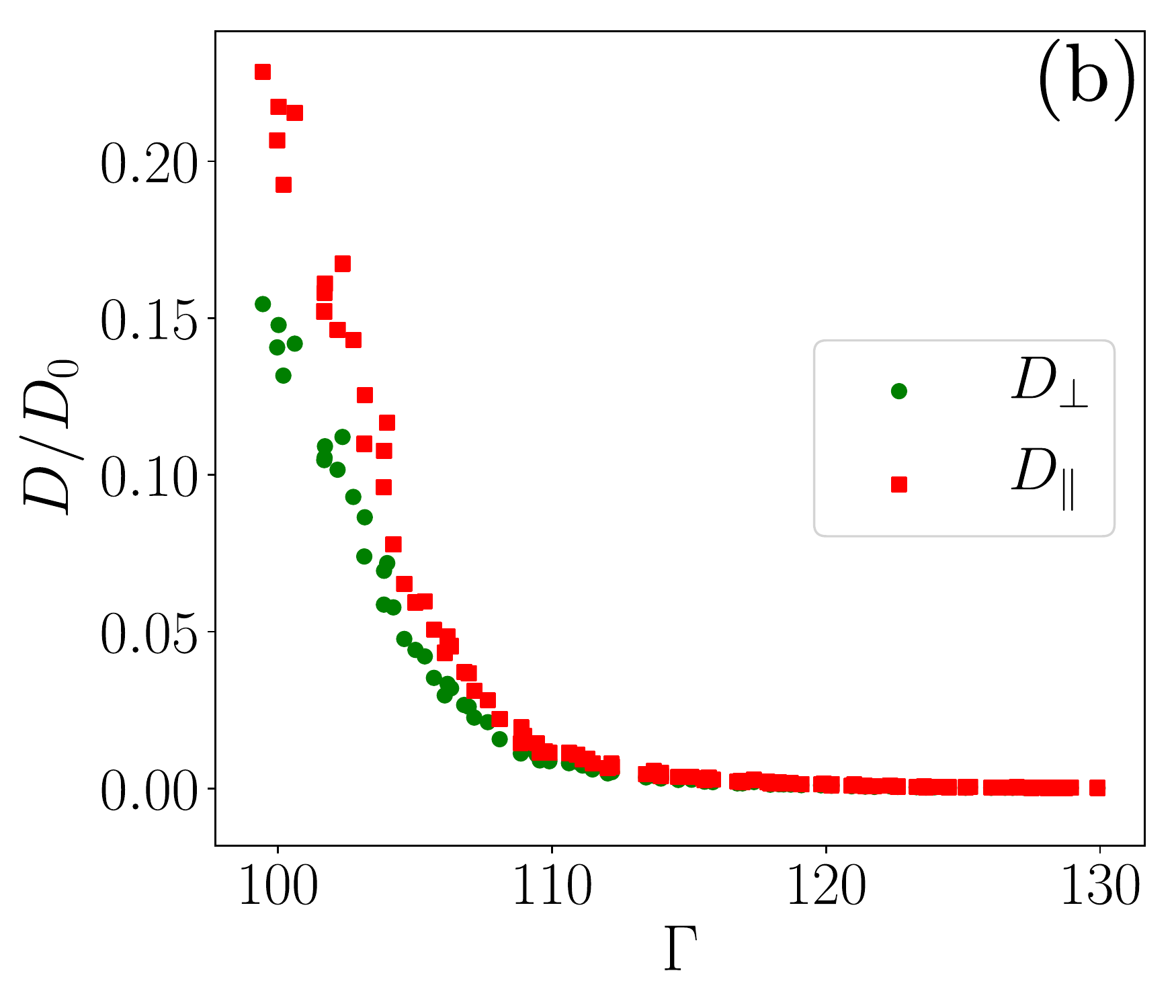}	
\caption{\small Diffusion constants $D_x\equiv D_\perp$ and $D_y\equiv D_\parallel$ vs $\Gamma$ in the maximally commensurate 1D periodic potential where the commensurability parameter (\ref{eq:commensurability_parameter}) is $p_c=1$. Panels (a) and (b) refer to the different strength of the potential,  $A/E_C=0.0157$ and $A/E_C=0.00157$, respectively. The simulations are done in the same way as in obtaining the data  in Fig.~\ref{fig:diff_isotropic}, the results refer to $N=1600$ electrons; both elastic and inelastic scattering are taken into account. For every point $\Gamma$, the system is prepared in a liquid state with $\Gamma =45$ and then cooled down in the periodic potential.}
	\label{fig:diffusion_commensurate}
\end{figure}

The data of simulations of the electron diffusion in the presence of a 1D maximally commensurate ($p_c=1$) periodic potential are shown in Fig.~\ref{fig:diffusion_commensurate}.  In the simulations, where we keep the electron density fixed, a natural scale for the potential is the Coulomb energy $E_C=e^2(\pi n_s)^{1/2}$, Eq.~(\ref{eq:Gamma}). The potential makes a very strong impact on the dynamics where it is still  much smaller than $E_C$. The vibration frequency $\nu_0$ is also small, $\nu_0^2/\omega_p^2 =4(\pi^3/3)^{1/2}A/E_C\ll 1$. To have a flavor of the numbers involved we note that, for the interelectron distance $a_s = 1~\mu$m, the amplitude of the potential in panels (a) and (b) corresponds to $A=0.5$~K and $A=0.05$~K.

The dramatic effect of a periodic potential on the diffusion and on the character of the liquid-to-solid transition is seen from the comparison of Figs.~\ref{fig:diffusion_commensurate} and \ref{fig:diff_isotropic}. Even for a very small potential in Fig.~\ref{fig:diffusion_commensurate}~(b) the diffusion coefficient approaches zero smoothly, indicating a continuous liquid to solid transition. The transition occurs around $\Gamma= 110$, the value significantly smaller (the temperature higher) than in a free electron system. 

For a 10 times stronger potential in Fig.~\ref{fig:diffusion_commensurate}~(a), the transition occurs for still smaller $\Gamma$. In contrast to the previous case, the coefficients of diffusion transverse and parallel to the potential troughs, $D_x\equiv D_\parallel$ and $D_y\equiv D_\perp$, are significantly different even very close to the transition, i.e., diffusion transverse to the troughs freezes out with decreasing temperature (increasing $\Gamma$) significantly earlier than diffusion along the troughs. A strong drop of $D_x$ compared to the value in the absence of the potential was seen in the simulations \cite{Kalia1983} for $\Gamma=36$ and the potential stronger than the one used here by a factor of $\sim 2$. 

Since close to the transition the electrons within a trough are strongly correlated, quasi-one-dimensional diffusion along the troughs comes either from density fluctuations (an extra electron or a missing electron in a small region) or has a purely numerical source: in a finite system a whole electron row of length $L_y$ can shift by an inter-electron distance, which would contribute to the diffusion, according to the way the diffusion is calculated. This contribution may lie behind the spread of the data on $D_y$ in Fig.~\ref{fig:diffusion_commensurate}~(a), see below. In contrast, nearly isotropic diffusion in Fig.~\ref{fig:diffusion_commensurate}~(b)  shows that  a weak potential  (small $A/k_BT$) weakly impedes the diffusion across the troughs compared to the diffusion along the troughs, as both are dominated by the disorder in the electron system. Nevertheless, even a weak potential changes the character of the transition between the liquid and solid phases.
\begin{figure}[h]
	\centering
	\		\includegraphics[width =4.2truecm]{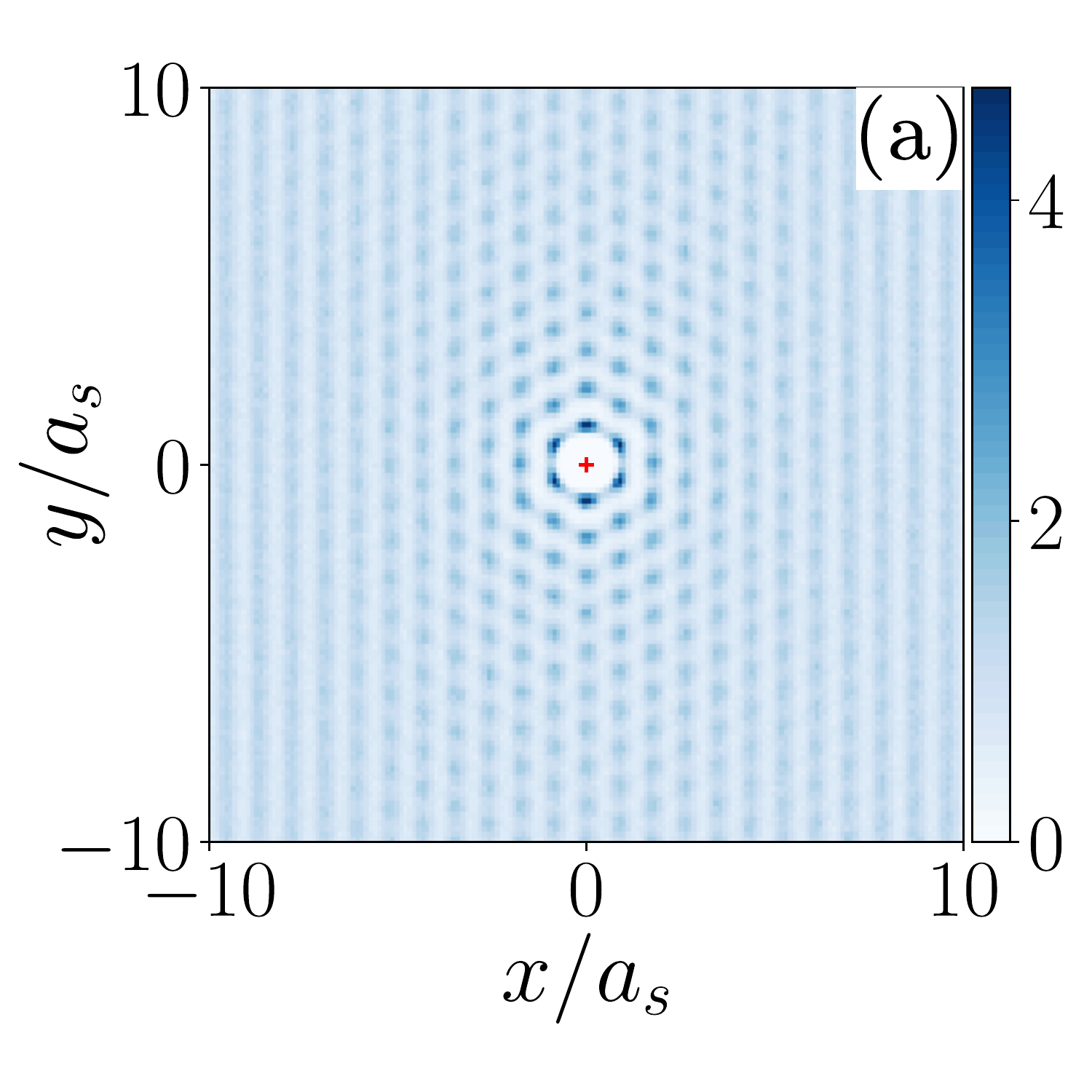} 
		\qquad	
			\includegraphics[width =4.2truecm]{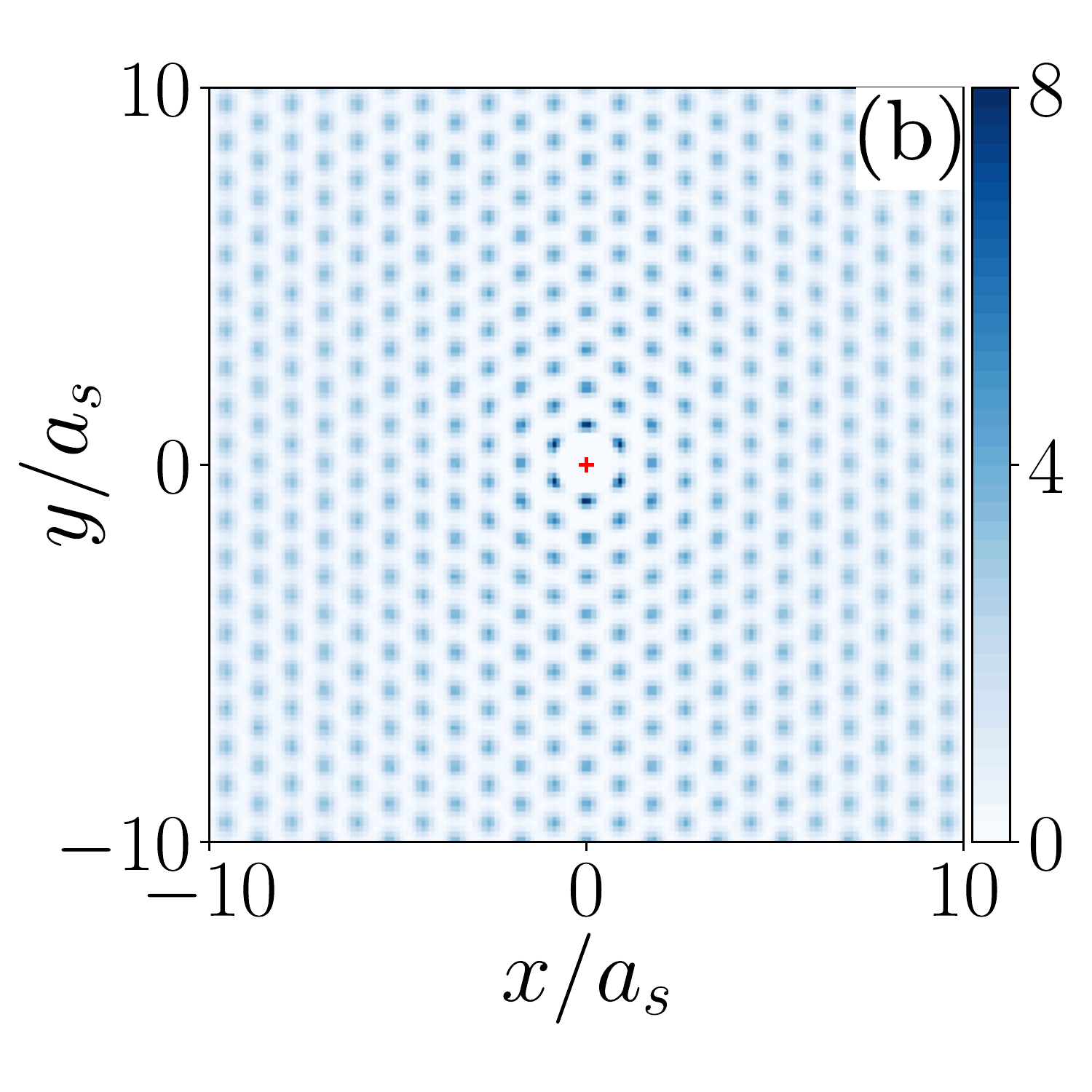} 
		\caption{\small Pair correlation function $g^{(2)}(\rb)$ of the electron system in the external potential for the same parameters as in Fig.~\ref{fig:diffusion_commensurate}~(b): $A/E_C=0.00157$ and $p_c=1$. \textbf{(a)} $\Gamma=100$, \textbf{(b)} $\Gamma=110$. The correlation function is dimensionless. The red cross indicates the location of the origin.}
	\label{fig:pair_correlator_commensurate}
\end{figure}

An insight into the change of the ordering imposed by a weak potential can be gained from Fig.~\ref{fig:pair_correlator_commensurate} that shows the pair correlation function 
\begin{align}
\label{eq:pair_correlation}
g^{(2)}(\rb) = \frac{1}{n_s N(N-1)}\sumprime{n,m}\delta[\rb - (\rb_n -\rb_m)].
\end{align} 
Figure~\ref{fig:pair_correlator_commensurate} refers to the weak periodic potential, the value of the amplitude $A$ is the same as in Fig.~\ref{fig:diffusion_commensurate}~(b). Even above the liquid to solid transition, as seen from Fig.~\ref{fig:pair_correlator_commensurate}~(a), there is partial electron ordering along the troughs of $U(x)$. As $\Gamma$ increases, the electrons become more localized within the troughs and there also increases the correlation between electrons in different troughs, ultimately resulting in the formation of a Wigner crystal.  

\section{Incommensurate periodic potential}
\label{sec:incommensurate}

\subsection{Strong incommensurability}

We found that, for strong incommensurability,  $p_c\lesssim 0.3$, a weak periodic potential ($A\ll E_C$) makes a small effect on the diffusion and on the position and the character of the crystallization transition. In particular, for $p_c=12/40$ and $A/E_C = 0.0157$ the dependence of the diffusion coefficients on $\Gamma$ is essentially indistinguishable, within the simulation error, from that for the free electron system in Fig.~\ref{fig:diff_isotropic}~(a). This is in qualitative difference with a system of noninteracting electrons, where diffusion transverse to the potential troughs is strongly modified once $A\gtrsim k_BT$, as the electrons have to overcome a high potential barrier. The density of the electrons with the appropriate energy is  $n_s(2\pi A/k_BT)^{1/2}\exp(-2A/k_BT)$ for $2A\gg k_BT$. It is much smaller than $n_s$. In contrast, the many-electron system averages out the potential where it is not very strong and is far from commensurability.

\subsection{Weak incommensurability}
\label{subsec:weak_incommensurability}

Where a periodic potential is close to being strongly commensurate, i.e., $p_c$ becomes close to 1, it makes a pronounced effect on the electron system even if it is weak. For systems with a short-range inter-particle coupling (for example, crystal monolayers deposited on a crystalline structure with different periodicity), this effect was considered in Refs.~\cite{Pokrovsky1979,Pokrovskii1980}, see also \cite{Bak1982}. It was predicted that there may be formed a lattice of solitons that would minimize the overall elastic energy. The long-range interaction and the associated inapplicability of the conventional elasticity theory can change the character of the crystallization transition for a Wigner crystal. 

\begin{figure}[h]
	\centering
	\includegraphics[width = 4.5truecm]{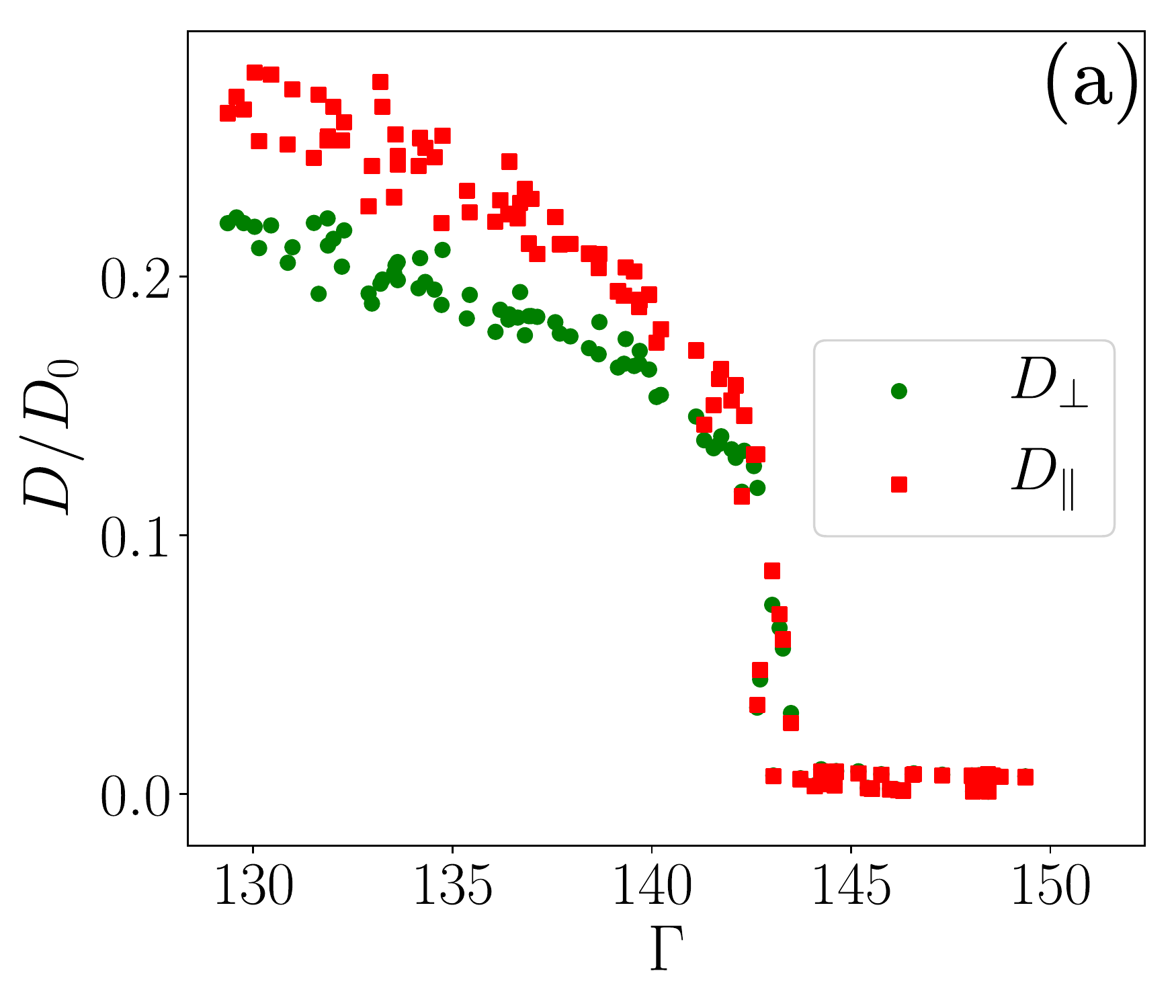}\qquad
\includegraphics[width = 4.5truecm]{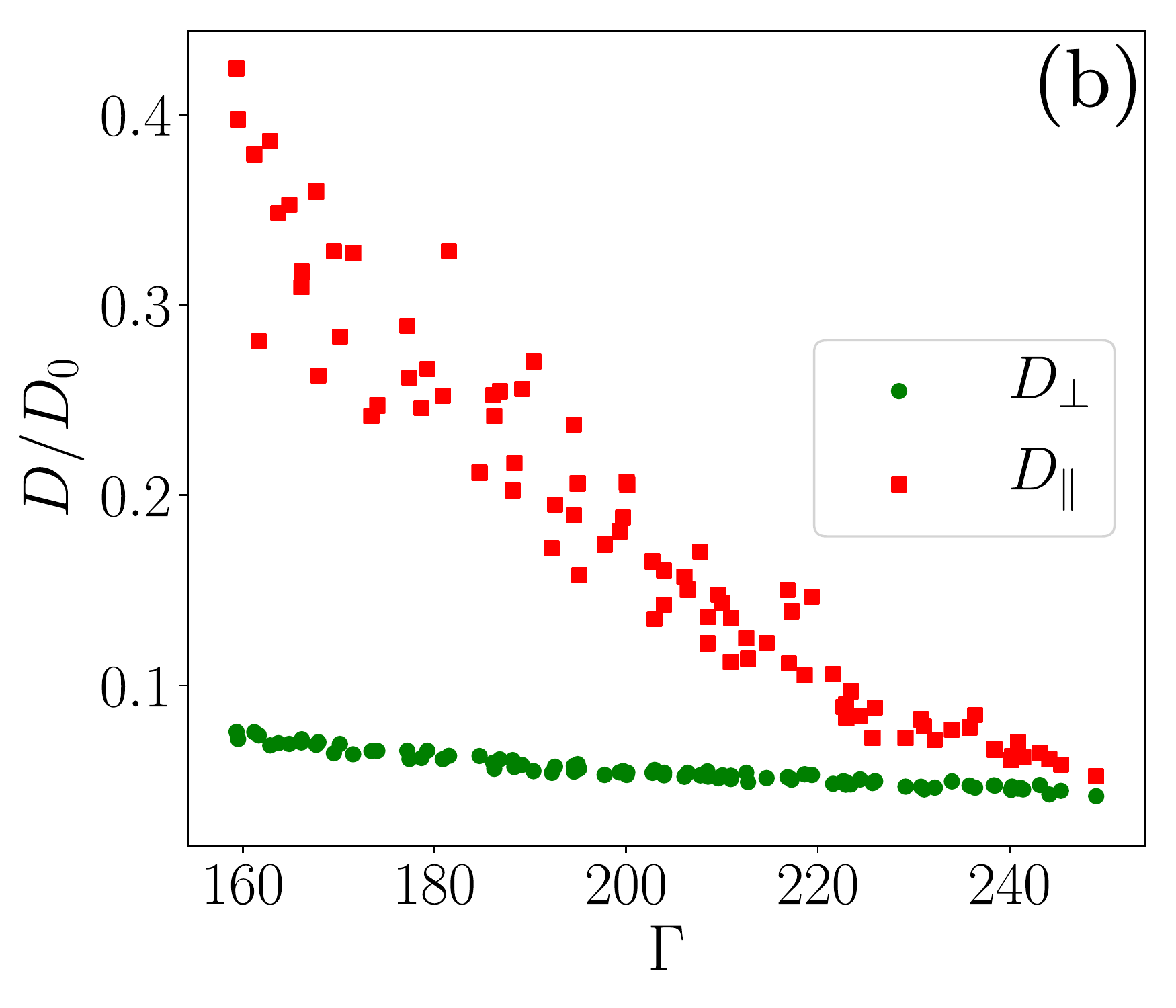}\qquad
\includegraphics[width = 4.5truecm]{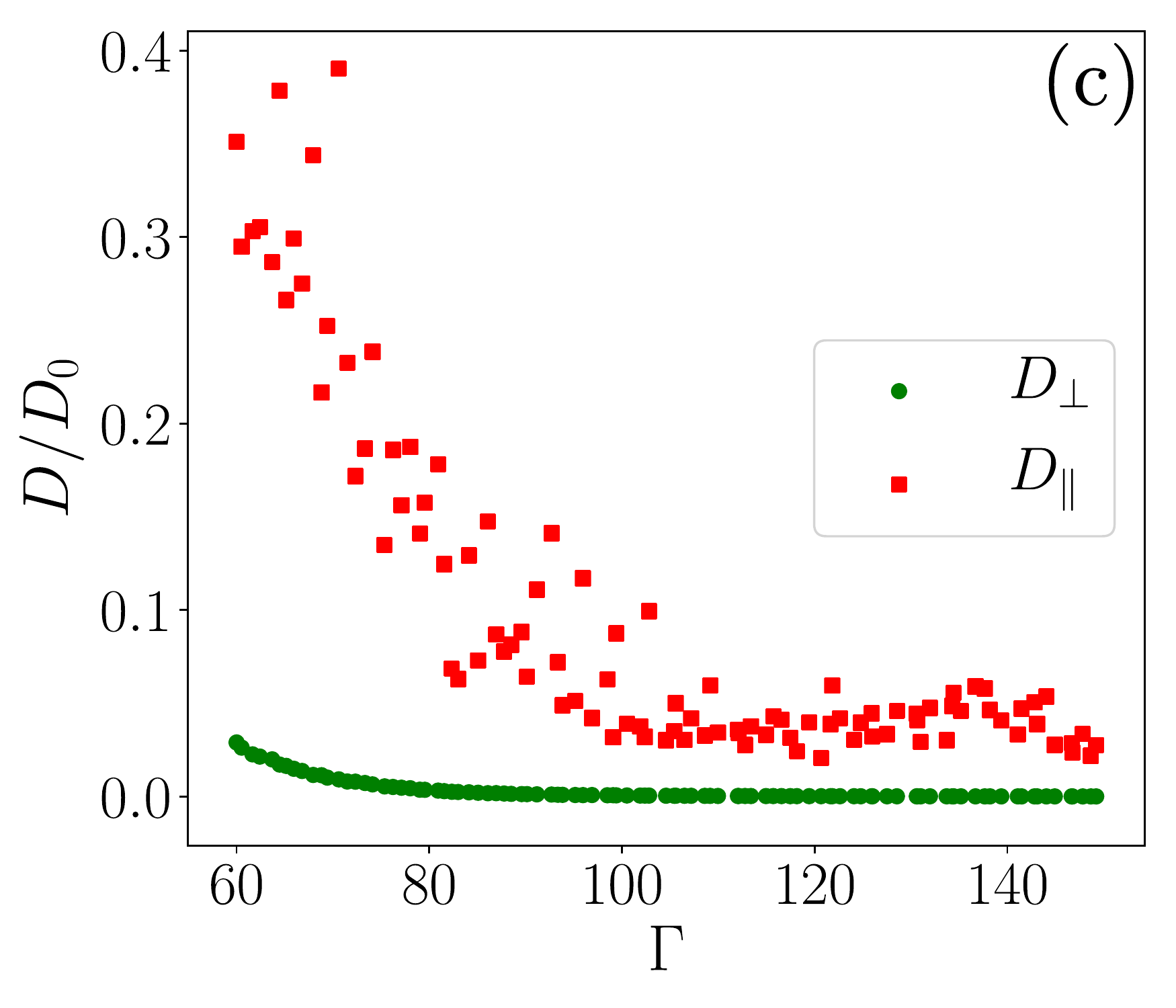}
	\caption{\small Diffusion constants $D_\perp$ and $D_\parallel$ vs $\Gamma$ in the periodic potential for $A/E_C=0.0157$ and  $N=1600$. The simulations include both elastic and inelastic scattering. For every point, the system is prepared as a liquid at $\Gamma=45$, cooled down to the temperature corresponding to a given $\Gamma$, and equilibrated for $10^6$ steps.  The commensurability parameter $p_c$ in (a), (b), and (c) is 30/40, 34/40, and 37/40, respectively. }
	\label{fig:diffusion_incommens}
\end{figure}

In Fig.~\ref{fig:diffusion_incommens} we show the dramatic effect of the weakly incommensurate potential on the electron diffusion.  The chosen potential height is the same as in Fig.~\ref{fig:diffusion_commensurate}~(a). We find that, not too close to the maximal  commensurability, $p_c=30/40$, the liquid to solid transition is seen as an abrupt change of the diffusion coefficient. This behavior is similar to that of the free electron system. However, the values of the diffusion coefficients are different and moreover, they can display strong anisotropy. A strong anisotropy is seen also for other values of $p_c$. Strikingly,  we found that for $p_c = 30/40$ and 32/40, the value of $\Gamma$ where the system ``freezes'' into a state with no diffusion 
increases with the increasing $p_c$ and is significantly larger than in the free system. 
%Yet closer to the strong commensurability, already for $p_c=37/40$, for the chosen potential strength, the critical value of $\Gamma$ decreases below  that for the free system, the coefficient of diffusion transverse to the potential troughs approaches zero smoothly with the increasing $\Gamma$, as in the commensurate case, and the discontinuity of the coefficient of diffusion along the troughs is within the simulation error. 

A part of the diffusion anisotropy can be attributed to the finite size of the system. Since the system is translationally invariant along the $y$-axis and is finite, whole rows of electrons can move along this axis, as mentioned previously. We reconstructed what happens to the electrons initially confined to the main region  $0\leq x < L_x, \; 0\leq y < L_y$ when we look at their positions in the extended region of $x,y$.  We saw that, for $p=34/40$, even for $\Gamma=200$ the electrons remain largely confined to the region $x\in (0, L_x)$, but along the $y$-axis the whole rows shift beyond the boundaries at $y=0, L_y$. We remind that, in the simulations, once an electron goes over the boundary of the main region, it is reinjected on the other side; however, in calculating the diffusion coefficient we take into account the total displacement, and therefore a displacement of an electron row gives a large contribution to the diffusion coefficient $D_y$. A displacement of a whole row is a rare event. Therefore even the large number of steps we are using is apparently not sufficient for obtaining a good statistics, as we see from the dependence of $\langle [y(t)-y(0)]^2\rangle$ on time. However, improving statistics in this case does not make sense, since we are dealing with a finite-size effect, which is of a limited physical interest.

The decrease of the freezing temperature (the increase of the critical $\Gamma$) as the system starts approaching the strong commensurability is a consequence of the competition between different structures that almost fit into the periodic potential. They correspond to the electron crystal being deformed and tilted with respect to the potential grating. If the difference in the energy densities of the deformed crystal and the electron liquid is smaller than in the free system, the freezing temperature should go down. 

\begin{figure}[h]
	\centering
\includegraphics[height =5.0truecm]{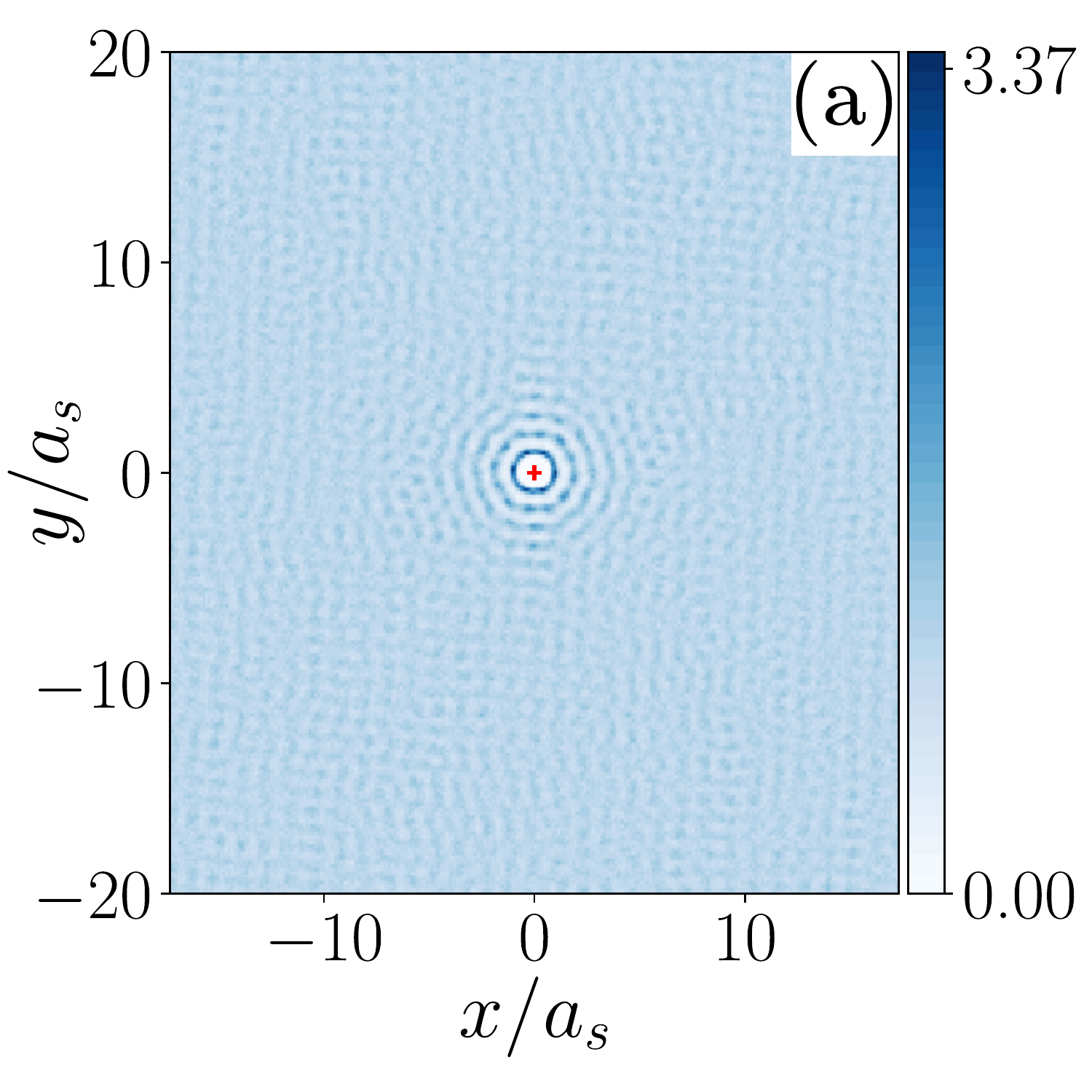}\quad
\includegraphics[height =5.0truecm]{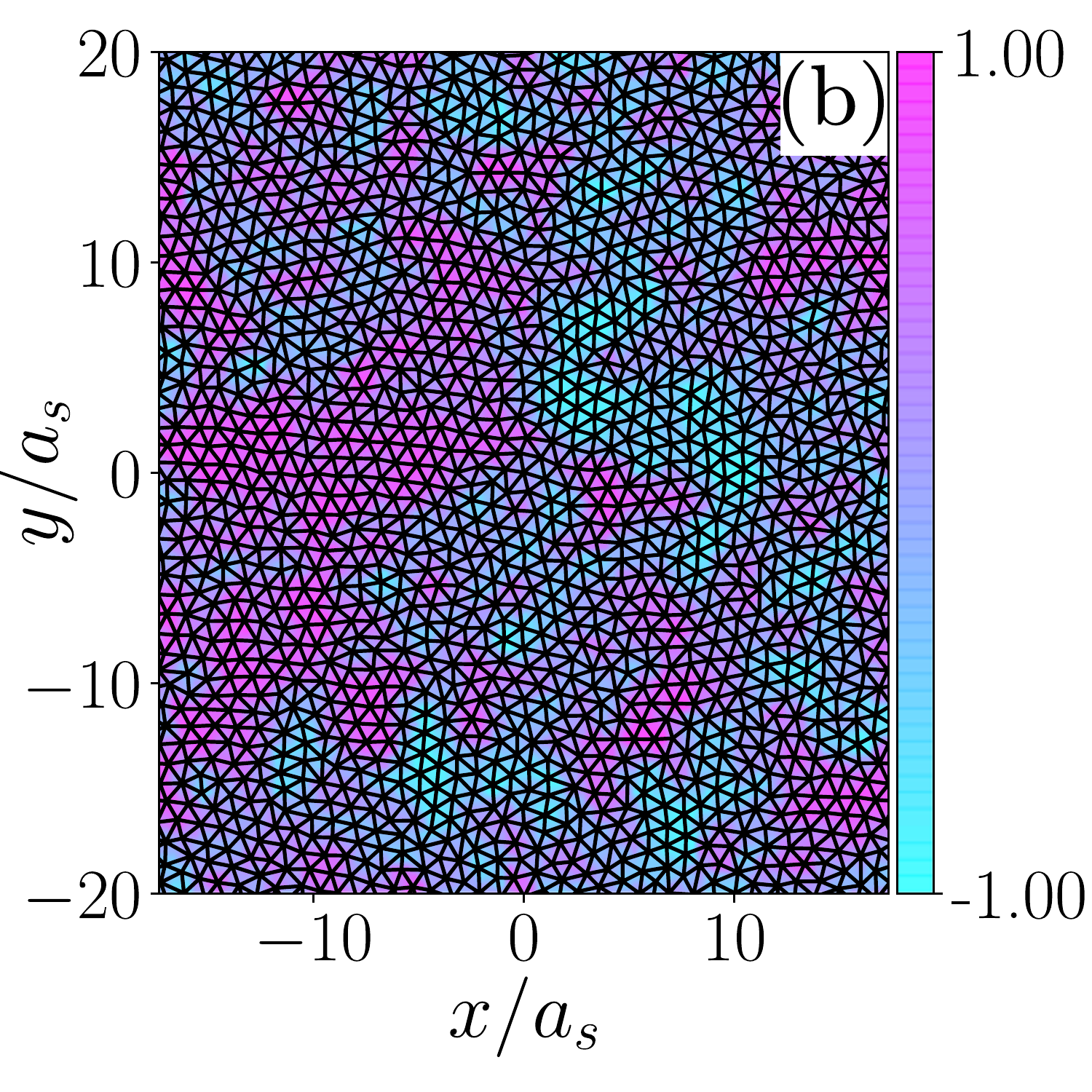}\quad		
\includegraphics[height =5.1truecm]{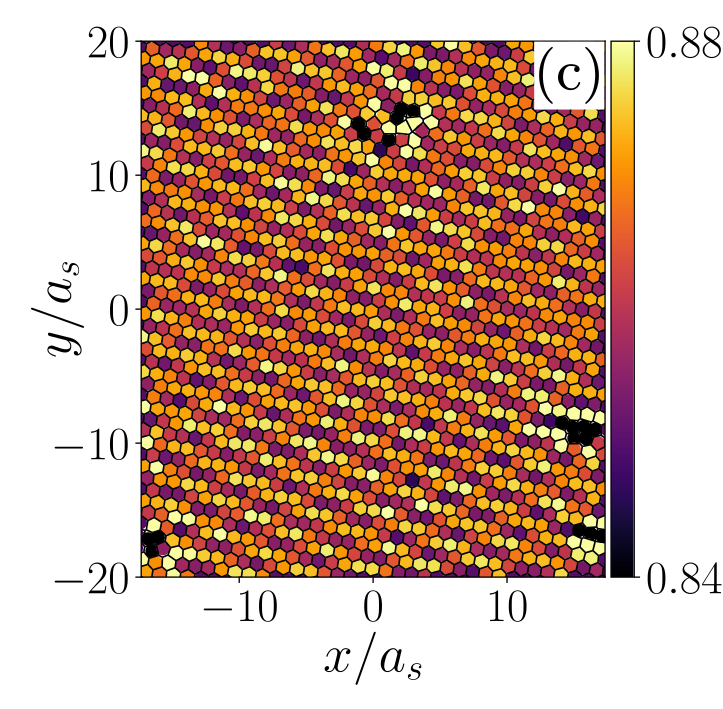} 
\caption{\small The results of the simulations of a system of $N=1600$ electrons in the 1D periodic potential for $A/E_C = 0.0157$ and the commensurability parameter $p_c=3/4$. (a) Pair correlation function $g^{(2)}(\rb)$, Eq.~(\ref{eq:pair_correlation}), $\Gamma=140.3$.  (b) Delaunay triangulation of a snapshot of the system with electrons represented by vertices of the triangulation,$\Gamma=140.3$. The background color shows $\mathrm{Re}\left[\psi_6^{(n)}\right]$, Eq.~(\ref{eq:hexatic}). (c)  Voronoi tessellation of the Wigner crystal for $\Gamma \approx 149$. Voronoi cells are colored depending on their surface area: from dark purple for smaller cells to light yellow for larger cells. The variation of the cell areas is from $0.97/n_s$ to $1.02/n_s$. The electron positions are time-averaged to account for small-amplitude thermal fluctuations. Clear rows of increased density appear at a nonzero angle with the modulation direction. They correspond to the incommensuration solitons \cite{Pokrovskii1980}.}
	\label{fig:solitons}
\end{figure}

As seen from the  Delaunay triangulation in Fig.~\ref{fig:solitons}~(b), for $\Gamma$ already above the critical value for the free system, the liquid phase in the modulated system consists of regions with the orientation largely correlated with the potential symmetry. The color in Fig.~\ref{fig:solitons}~(b) shows  the real part of the bond orientational order parameter $\psi_6^{(n)}$, 
\begin{align}
\label{eq:hexatic}
\psi_6^{(n)}=M_{\rm nn}^{-1}\sum_m\exp(6i\theta_{mn}),
\end{align}
where the sum runs over the nearest neighbors of the $n$th electrons, $M_{nn}$ is the number of the neighbors, and $\theta_{nm}$ is the angle between the bond to neighbor $m$ and the $x$-axis. By construction,  $\psi^{(n)}_6$ is constant within a Voronoi cell centered at an $n$th electron. In Fig.~\ref{fig:solitons}~(b), different regions in the electron liquid are aligned close to the $x$ axis (magenta) or close to the $y$-axis (blue). The sizes of these regions are different and vary from a snapshot to a snapshot. The electrons diffuse along the boundaries and the boundaries themselves move. The pair correlation function in Fig.~\ref{fig:solitons}~(a) shows a short-range order for $\Gamma\approx 140$ with weak overall density modulation imposed by the potential. 

In the crystalline phase, for several values of $p_c$ we observed the onset of periodically repeated ranges of increased density, the incommensuration solitons \cite{Pokrovskii1980}. They are shown in Fig.~\ref{fig:solitons} for $p_c=3/4$. The data are obtained by cooling the system from the liquid state at $\Gamma=45$ to the targeted value of $\Gamma$. We found that the resulting solitonic structure is metastable, it depends on $\Gamma$. The observation of the solitons shows that the onset of solitons is robust. It is not limited to systems with a short-range interaction, but rather applies to systems with the Coulomb coupling, even though the Wigner crystallization in such systems in the presence of a periodic potential is qualitatively different from that in systems with a short-range coupling. 

\begin{figure}[h]
\centering
\includegraphics[height=5.0truecm]{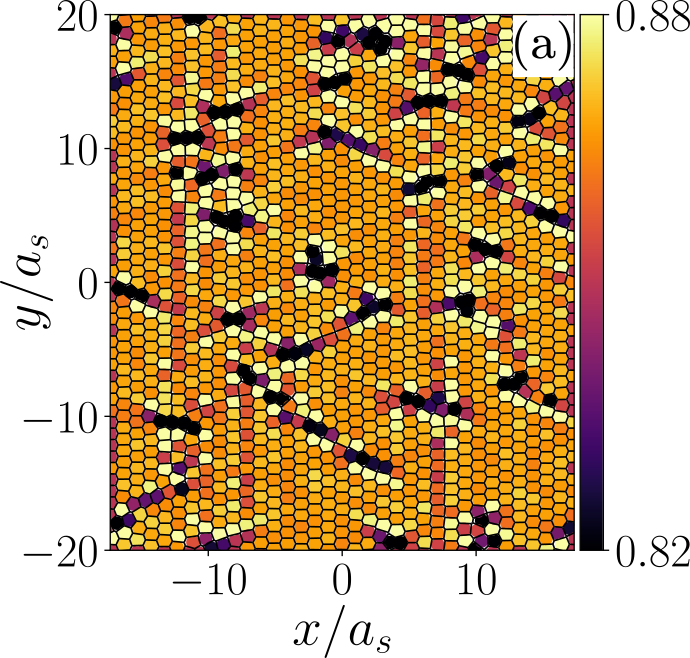}\qquad
\includegraphics[height=5.0truecm]{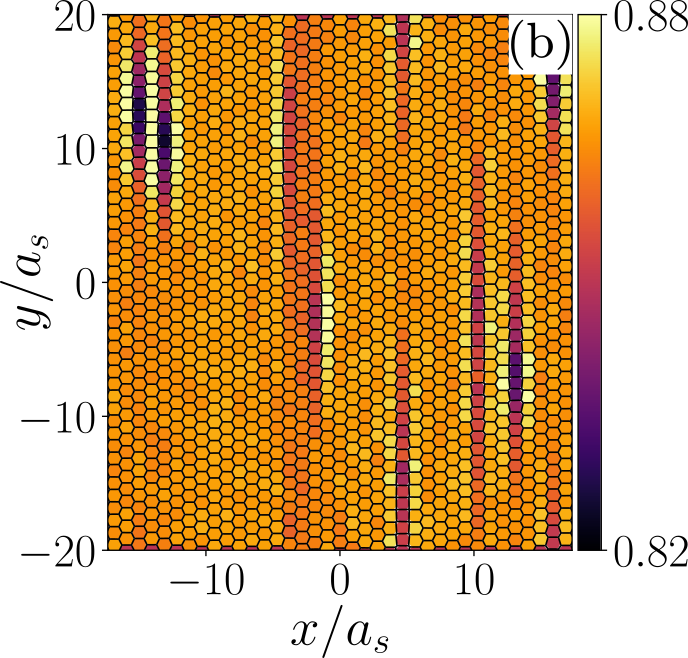}
\caption{\small Voronoi cells for N=1600 electrons in the 1D periodic potential for $A/E_C=0.0157$; (a) $p_c=34/40$, $\Gamma=240$, and (b) $p_c=37/40, \Gamma= 141$. The coloring scheme is the same as in Fig.~\ref{fig:solitons}(c), and the data are obtained by cooling the system from the liquid state at $\Gamma=45$.}
\label{fig:strong_incommensurability}
\end{figure}

The structure of the system and the dynamics change dramatically as $p_c$ further approaches the maximal commensurabiltiy value $p_c=1$. The Voronoi tessellation in Fig.~\ref{fig:strong_incommensurability}(a) shows that, for $p_c=34/40$, the system remains disordered even for a very large $\Gamma$. Compared to a Wigner crystal, it has a large number of defects that stretch in certain symmetry directions. Such defects can move. This should be the reason of the persistent diffusion in  Fig.~\ref{fig:diffusion_incommens}(b). Interestingly, as $\Gamma$ increases, the coefficients of diffusion along and transverse to the troughs become essentially equal. For $p_c=34/40$, this happens for $\Gamma\approx 240$. We noticed that the motion of electron rows as a whole along the potential troughs is strongly suppressed for large $\Gamma$ and diffusion becomes almost isotropic. This is a natural consequence of the strong correlations in the system. Interestingly, we did not see freezing of the diffusion for $p_c=34/40$ even for $\Gamma=240$. In contrast, 

For $p_c=37/40$ motion transverse to the troughs is frozen already for $\Gamma\approx 90$, whereas motion along the troughs does not freeze even for $\Gamma=220$. It is fully dominated by the random motion of the rows, and therefore in Fig.~\ref{fig:diffusion_incommens}~(c) we show data only for $\Gamma<150$. 
Figure~\ref{fig:strong_incommensurability}(b) shows that for $p_c=37/40$ defects tend to cluster along the troughs into the regions of higher electron density surrounded in the neighboring troughs by regions of lower density. For the values of $\Gamma$ in Figs.~\ref{fig:strong_incommensurability}(a) and (b), the defects already have a pronounced orientational structure. Our results do not show a long-range order in the system for $p_c=34/40$ or 37/40. It might be related to the system being too small. One expects that for  $p_c\to 1$ the  behavior of the infinite system should approach that for the maximal commensurability.

\section{Conclusions}
\label{sec:conclusions}

This paper reports the results of the molecular dynamics simulations of the electron system on liquid helium. We concentrated on electron self-diffusion  in this strongly correlated system, as it characterizes the long-term electron dynamics. We have found that, where the electrons are in the liquid phase, the diffusion is ``normal'', with the mean square displacement  proportional to time. This is the case both for a uniform electron system and for an electron system placed into a 1D periodic potential, although the values of the diffusion coefficients can be very strongly changed by the potential even if the modulation of the electron potential energy is much weaker than the characteristic energy of the electron-electron interaction. 

Important distinctions of our simulations of the electron dynamics from the previous work are not only the large size of the system but also that we take into account both elastic and inelastic electron scattering off the excitations in the thermal reservoir. We model the scattering as short events with the rates that correspond to the scattering rates for electrons on the helium surface. The scattering rate is small compared to the plasma frequency and $k_BT/\hbar$, so that the electron motion is underdamped. Yet the scattering leads to the relaxation of the total electron momentum and, even more importantly in the context of this paper, it leads to thermalization of the electron system. This is particularly helpful in the analysis of the dynamics in the vicinity of the liquid to solid transition, where the thermalization of an isolated electron system slows down. Another important feature of our simulations is that we use over $10^7$ time steps. We found that such a long time is necessary to achieve full equilibration and improve the statistics.

For a free electron system, the diffusion coefficient monotonically decreases with the decreasing temperature. At the liquid to solid transition, it drops to zero in an extremely narrow temperature range, essentially discontinuously. In our long simulations we are unable to see a hysteresis; this should be contrasted with the observation of the hysteresis in the earlier work \cite{Kalia1981} where the number of steps was smaller by a factor $\sim 10^3$ (we saw hysteresis when we used a smaller number of steps). For the long simulations we use, within the simulation error, the sharp change of the diffusion coefficient occurs for the same $\Gamma$ whether we cool the system down starting from a random electron configuration or heat it up starting from a Wigner crystal.

A maximally commensurate 1D periodic potential dramatically changes the electron dynamics. We see this change for the ratio of the amplitude of the modulation of the potential energy $A$ to the energy of the electron-electron interaction $E_C$ as small as $\sim 1.6\times 10^{-3}$. We show that, in contrast to systems with a short-range coupling, in a 2D Wigner crystal placed into a commensurate 1D potential the mean-square electron displacement from a lattice site does not diverge. This indicates stability of the Wigner crystal and leads to a qualitative change of the character of the liquid to solid transition. We find that, in contrast to the free system, the diffusion coefficient smoothly goes to zero with the decreasing temperature. It becomes equal to zero, withing the accuracy of the data, for the value of the plasma parameter $\Gamma$ noticeably smaller than in the absence of the potential, $\Gamma\approx 110$ instead of $\Gamma\approx 137$. The critical value of $\Gamma$ strongly depends on the ratio $A/E_C$ and drops down to $\approx 75$ for $A/E_C\approx 0.016$. 

In the liquid phase, the diffusion in a 1D periodic potential is anisotropic. It is slower transverse to the potential troughs. The anisotropy strongly depends on the strength of the potential and the commensurability. Unexpectedly, the critical value of $\Gamma$ where the diffusion coefficient turns to zero depends on the commensurability parameter nonmonotonically. Even for a comparatively weak potential, $A/E_C\approx 0.016, $ as the period of the potential is changed toward the maximal commensurability, the critical $\Gamma$ first becomes higher than for the free-electron system and thus very much higher than for the maximally commensurate potential. In our finite system with periodic boundary conditions, for $A/E_C\approx 0.016$  we did not see the decrease of the critical $\Gamma$ expected in the region of the crossover to the behavior in the maximally commensurate potential. This crossover and its dependence on the strength of the potential warrant further investigation. 

Not too close to the maximal commensurability, we have seen solitons in the crystalline phase. They appear even though the longitudinal long-wavelength excitations in the Wigner crystal qualitatively differ from acoustic phonons. 

In contrast to colloidal particles, where motion of individual particles can be traced in the experiment, cf. \cite{Chowdhury1985,Wei1998,Mangold2003,Lu2013,Li2016}, there are no established means to trace individual electrons on the helium surface. A potentially feasible  experimental approach to characterizing the electron motion is to study the decay of the density correlation function. The measurement can be done if one places a periodic  one-dimensional array of nanoscale sensors of the local potential beneath the helium surface. This array can be placed on the same substrate as the array of the electrodes that create the periodic potential. Fluctuations of the electron density will lead to fluctuations $\delta {\cal Q}$ of the total charge of the array. In the presence of self-diffusion, one may expect that the correlator $\langle \delta {\cal Q}(t) \;\delta {\cal Q}(0)\rangle$ will decay as $\exp[-(2\pi/a_m)^2D_m t/2]$, where $a_m$ is the inter-sensor distance and $D_m$ is the component of the diffusion coefficient along the sensor array.

Electrons on helium can be used also to observe the lattice of solitons that emerge where electrons crystallize in a periodic potential and the system is close to the maximal commensurability. A convenient way of detecting solitons is by measuring the electron transport along the potential troughs. Because of the weak coupling to the helium surface, the current-voltage characteristic becomes nonlinear already for a weak driving field in the presence of a translational order in the electron system. We expect that the characteristic current is determined by the velocity of the surface waves on helium with the wave vector equal to the vector of the reciprocal soliton lattice. The underlying Bragg-Cherenkov mechanism \cite{Dykman1997b,Vinen1999} has been revealed for a Wigner crystal and used to establish Wigner crystallization in the absence of an extra periodic potential, cf.~\cite{Kristensen1996, Glasson2001, Rees2016,Rees2017}. Since the lattice constant of the solitons is significantly larger than the inter-electron distance, the phase velocity of the helium surface waves is smaller than in the case of the Wigner crystal.  Respectively, the nonlinearity of the current-voltage characteristic should emerge for significantly smaller current than that in the absence of a periodic potential and should directly reveal the soliton spacing.

\noindent
{\bf Acknowledgments}

We are grateful for the discussion of the results of this paper to the participants of The International Workshop on Electrons and Ions in
Quantum Fluids and Solids (Japan 2018) and the organizer of this workshop K. Kono. This research was supported in part by the NSF-DMR Grant 1708331 
%%%%%%%%%%%%%%%%%%%%%%%%%%%%%%%%%%%%%%%%%%%%%%%%%%%%%%%%%%%%

\hfill

\noindent
{\large\bf Appendix}

\appendix
\section{Numerical integration of the equations of motion}
\label{sec:appA}
\numberwithin{equation}{section}

\setcounter{equation}{0}

To study properties of the electrons on helium we integrate classical equations of motion numerically. We use HOOMD-Blue \cite{Anderson2008,Glaser2015} as the base code for our simulations, with the integrator, interaction potentials, and external forces developed specifically for our study\footnote{Our fork of the HOOMD-Blue code can be found at \url{https://github.com/kmoskovtsev/HOOMD-Blue-fork}. The scripts used in the simulations are available at \url{https://github.com/kmoskovtsev/Electrons-on-Helium-Scripts}}. The code is designed to be used on a graphics processing unit (GPU) which gives up to sixteen-fold acceleration compared with a single CPU core.  For integration, we use the standard velocity Verlet algorithm, which is symplectic \cite{Hairer2006}. The algorithm is implemented numerically as follows:
	\begin{align}
	\vb_i\left(t + \Delta t/2\right) &= \vb_i(t) + \frac{1}{2}\ab_i(t)\Delta t,\nonumber \\
	\rb_i\left(t + \Delta t\right) &= \rb_i(t) + \vb_i\left(t + \Delta t/2\right) \Delta t,\nonumber \\
	\vb_i(t + \Delta t) &= \vb_i\left(t + \Delta t/2\right) + \frac{1}{2}\ab_i(t + \Delta t)\Delta t,
	\end{align}
where $\vb_i$ is the $i$th particle velocity, and the acceleration $\ab_i(t)$ is evaluated based on forces derived from the positions $\rb_j(t)$ of all particles at time $t$. In the calculation, time is discretized and $\Delta t$ is the discretization step.

We modified the algorithm to incorporate discrete scattering events, which are described by an abrupt random change of the electron velocity $\vb \rightarrow \vb'$, or the corresponding change of the electron momentum $\hbar\kb\to \hbar(\kb+\Delta \kb)$, that obeys a certain probability distribution. As mentioned in Sec.~\ref{sec:integrating}, in the classical description, electron positions do not change in scattering events, and therefore the electron potential energy is not changed either. This is automatically  satisfied in the calculation if the velocity is changed at integer time points. Importantly, in elastic scattering, which is the dominating scattering mechanism, $|\vb(t) + \Delta\vb| = |\vb(t)|$, where $\Delta\vb$ is the velocity change at time $t$, and then the total energy is conserved to the first order in $\Delta t$. Indeed, 
	\begin{align}
	\vb(t+\Delta t) &= \vb(t) + \Delta\vb + \frac{1}{2}(\ab(t) + \ab(t+\Delta t))\Delta t,\nonumber \\
	\rb(t + \Delta t) &= \rb(t) + \Delta t\left[\vb(t) + \Delta\vb + \frac{1}{2}\ab(t)\Delta t \right].
	\end{align}
Then it is easy to see that the energy difference over one step vanishes in the first order in $\Delta t$,
	\begin{align}
	E(t + \Delta t) - E(t) = \frac{m}{2}\left(\vb^2(t + \Delta t) - \vb^2(t)\right) + U\left(\rb(t + \Delta t)\right) - U\left(\rb(t)\right) = O(\Delta t^2).
	\end{align}
Since the scattering events in our simulations are rare, about one scattering event per $10^5$ time steps per particle, this accuracy is sufficient for our purpose.

We have chosen $\Delta t \approx (2\pi/\omega_p)/50$ for most of the simulations, where $\omega_p$ is the short-wavelength plasma frequency given by Eq.~(\ref{eq:classical_dynamics}). Reducing $\Delta t$ proved to have no visible effect on the studied phenomena.

As described in Sec.~\ref{sec:integrating}, scattering processes are characterized by their rates $w(\kb\to\kb')$, which are calculated separately for the ripplon and phonon scattering. Different scattering events are uncorrelated. The scattering by ripplons is effectively elastic. Therefore, its rate $w^{\rm (rp)}(\kb\to\kb')$ depends only on the absolute value of $\kb$ and the angle $\theta$ between $\kb$ and $\kb'$. We tabulate the integrated over $\theta$ scattering rate for a given magnitude $k =|\kb|$ of  $\kb$ on a  grid of $200$ points. This rate is used to determine the probability to scatter at every time step. We then tabulate the inverse cumulative distribution of the scattering angles $\theta$ on $400$ points and use the inverse transform sampling  to generate a random value of $\theta$ if a scattering event occurs at a give time step.

In inelastic scattering, both $k'$ and $\theta$ must be determined for a scattering event. As Fig.~\ref{fig:wkkp_phonon} shows, the scattering rate varies quickly with $k'\equiv |\kb'|$ and rather smoothly with $\theta$ for each value of $k$. Therefore, it is convenient to implement the following radial scheme of generating $\kb'$. For each value of $k$, we compute the probability distribution of $k'$ by integrating the scattering rate over the scattering angle $\theta$. If a scattering event occurs at a given time step, we first generate the absolute value $k'$ based on this distribution. Then for every pair $k, k'$  we compute the probability distribution of  $\theta$. We draw a random $\theta$ using again the inverse transform sampling after $k'$ is generated in the previous step. The distributions are computed and tabulated in two arrays with sizes $200\times200$ and $200\times200\times10$. In addition to these distributions, we also tabulate the total rate of scattering from a state with wave vector $\kb$ into any other state to find the overall probability to scatter at each time step, as in the case of the ripplon scattering. For both scattering mechanisms, we use interpolation schemes to obtain continuous scattering distributions from the tabulated values.

An important test of the developed scheme is the stationary electron distribution function. The distribution over the electron momentum should be of the Maxwell-Boltzmann form with the temperature of the helium excitations. We checked that this is indeed the case both for noninteracting electrons and in the presence of the electron-electron interaction. 

%\bibliographystyle{bibl_try}
%\bibliography{c:/Users/Mark/Dropbox/Refs/md10test,c:/Users/mark/Dropbox/Kirill_e_ph/refs/electrons_on_He}

\end{document}